\documentclass[preprint,10pt]{aastex}

\usepackage{graphicx,amsmath,amssymb,txfonts,natbib}

\newcommand{\be}{\begin{equation}}
\newcommand{\ee}{\end{equation}}
\newcommand{\bdm}{\begin{displaymath}}
\newcommand{\edm}{\end{displaymath}}
\newcommand{\bea}{\begin{eqnarray}}
\newcommand{\eea}{\end{eqnarray}}
\newcommand{\beas}{\begin{eqnarray*}}
\newcommand{\eeas}{\end{eqnarray*}}

\newcommand{\av}[1]{\left< #1\right>}
\newcommand{\dk}{\mathrm{d}k}
\newcommand{\dq}{\mathrm{d}q}

\newcommand{\thetab}{\overline{\theta}}
\newcommand{\phib}{\overline{\phi}}
\newcommand{\betab}{\overline{\beta}}
\newcommand{\gammab}{\overline{\gamma}}
\newcommand{\mub}{\overline{\mu}}

\newcommand{\dcubedk}{\mathrm{d}^3\mathbf{k}}
\newcommand{\dphib}{\mathrm{d}\phib}
\newcommand{\dalpha}{\mathrm{d}\alpha}
\newcommand{\dz}{\mathrm{d}z}
\newcommand{\dy}{\mathrm{d}y}
\newcommand{\da}{\mathrm{d}a}

\newcommand{\alphaks}{\tilde{\alpha}_k}

\newcommand{\Pow}[1]{\mathcal{P}\left(#1\right)}

\newcommand{\FTrTrTr}{\mathcal{F}_{\tau\tau\tau}}
\newcommand{\FTsTsTs}{\mathcal{F}_{\tau_S\tau_S\tau_S}}
\newcommand{\FTtTtTt}{\mathcal{F}_{\tau_T\tau_T\tau_T}}
\newcommand{\BTrTrTr}{\mathcal{B}_{\tau\tau\tau}}
\newcommand{\BTsTsTs}{\mathcal{B}_{\tau_S\tau_S\tau_S}}
\newcommand{\BTtTtTt}{\mathcal{B}_{\tau_T\tau_T\tau_T}}

\newcommand{\BTrTrTs}{\mathcal{B}_{\tau\tau\tau_S}}
\newcommand{\BTrTsTs}{\mathcal{B}_{\tau\tau_S\tau_S}}
\newcommand{\BTrTvTv}{\mathcal{B}_{\tau\tau_V\tau_V}}
\newcommand{\BTsTvTv}{\mathcal{B}_{\tau_S\tau_V\tau_V}}
\newcommand{\BTrTtTt}{\mathcal{B}_{\tau\tau_T\tau_T}}
\newcommand{\BTsTtTt}{\mathcal{B}_{\tau\tau_T\tau_T}}
\newcommand{\BTvTtTv}{\mathcal{B}_{\tau_V\tau_T\tau_V}}

\newcommand{\khv}{\hat{\mathbf{k}}}
\newcommand{\phv}{\hat{\mathbf{p}}}
\newcommand{\qhv}{\hat{\mathbf{q}}}

\begin{document}

\title{Intrinsic Bispectra of Cosmic Magnetic Fields}

\author{Iain A. Brown}
\affil{Institute of Theoretical Astrophysics, University of Oslo}
\affil{0315 Oslo, Norway}
\email{i.a.brown@astro.uio.no}

\date{\today}

\begin{abstract}
Forthcoming datasets from the Planck experiment and others are in a position to probe the CMB non-Gaussianity with higher accuracy than has yet been possible and potentially open a new window into the physics of the very early universe. However, a signal need not necessarily be inflationary in origin, and possible contaminants should be examined in detail. One such is provided by early universe magnetic fields, which can be produced by a variety of models including during an inflationary phase, at phase transitions, or seeded by cosmic defects. Should such fields have been extent in the early universe they provide a natural source of CMB non-Gaussianity.

Knowledge of the CMB angular bispectrum requires the complete Fourier-space (or ``intrinsic'') bispectrum. In this paper I consider in detail the intrinsic bispectra of an early-universe magnetic field for a range of power-law magnetic spectra.
\end{abstract}

\maketitle

\section{Introduction}
Observations of the cosmic microwave background (CMB) strongly suggest that the dominant source of perturbations in the early universe was adiabatic and nearly Gaussian in nature \citep{Komatsu:2010fb}, consistent with an inflationary scenario. Inflationary scenarios also induce a small amount of non-Gaussianity, typically characterised by a parameter $f_{\mathrm{NL}}$ \citep[for example]{Liguori:2010hx} which from the WMAP seven-year results is currently constrained to be $-10<f_{\mathrm{NL}}<74$ at 95$\%$ confidence \citep{Komatsu:2010fb}. It is expected that these bounds will tighten significantly with the forthcoming Planck data \citep{Liguori:2010hx}. A detection of a significant $f_{\mathrm{NL}}$ could provide vital information about the early universe and high-energy physics, as it could rule out a variety of inflationary models including the simplest single-field models.

However, the inflationary model does not exclude the possibility that non-linear sources might also play a role in sourcing perturbations. Frequently-studied examples of such sources include cosmic defects \citep[for example]{Turok:1997gj} and cosmological magnetic fields \citep[for example]{Giovannini:2003yn,Tsagas:2004kv}.

Magnetic fields are observed on many scales in the cosmos, including on cluster scales with a coherence length on the order of megaparsecs and field strengths on the order of between nano-Gauss and micro-Gauss \citep{Kronberg:1993vk,Grasso:2000wj,Giovannini:2003yn,Xu:2005rb,Kahniashvili:2010wm}. Fields on larger scales are notoriously difficult to detect but there are some suggestions that such fields might exist with field strengths up to the order of micro-Gauss. Recent observations of radiation from blazars implies a lower limit in the extra-galactic medium of $B\gtrsim\mathcal{O}(10^{-16})$-$\mathcal{O}(10^{-15})$, permeating extracluster voids \citep{Neronov:1900zz,Tavecchio:2010ja,Dolag:2010ni}. The presence of fields on such fields implies either a primordial origin or an efficient transfer of fields from within galaxies deep into the intergalactic medium. The precise origin of fields on such large scales remains debated but magnetogenesis scenarios exist which were viable in the extremely early universe. Such scenarios can be produced directly during inflation \citep{Turner:1987bw,Bassett:2000aw,Prokopec:2004au,Giovannini:2007rh,Campanelli:2007cg,Bamba:2008my} or at a phase transition \citep{Baym:1995fk,Martin:1995su,Hindmarsh:1997tj,Boyanovsky:2002wa,Kahniashvili:2009qi}. Fields can also be produced by the production of non-linear vorticity from linear density perturbations\citep{Gopal:2004ut,Matarrese:2004kq,Takahashi:2005nd,Ichiki:2006cd,Siegel:2006px,Kobayashi:2007wd,Maeda:2008dv}, but the impact these have on the CMB is complicated by their evolving, non-trivial nature.

The impact primordial magnetic fields have on the CMB and its anisotropies have been well-studied, with \citep{Barrow:1997mj,Subramanian:1998fn,Durrer:1998ya,Koh:2000qw,Kahniashvili:2000vm,Mack:2001gc,Clarkson:2002dd,Lewis:2004ef,Kahniashvili:2006hy,Kahniashvili:2008hx,Yamazaki:2008gr,Finelli:2008xh,Paoletti:2008ck,Bonvin:2010nr,Giovannini:2009fu,Yamazaki:2010nf,Paoletti:2010rx,Kahniashvili:2010wm} being some instructive examples. While older literature tended to assume a homogeneous background component with an inhomogeneous perturbation, more recent work has typically focused on tangled configurations without a background component and I assume this throughout. These studies fairly consistently suggest that the field is constrained to be of at most nano-Gauss in magnitude. The spectral index is restricted to be approximately scale-invariant \citep{Yamazaki:2010nf,Paoletti:2010rx}, with limits growing extremely tight for a primordial magnetic field with index far from scale-invariance \citep{Caprini:2001nb}. A large-scale homogeneous field also introduces characteristic correlations between multipole moments with $\Delta l\in\{-2,0,2\}$ and $\Delta m\in\{0,\pm 1,\pm2\}$ which vanish in the standard scenario \citep{Kahniashvili:2008sh}.

However, the magnetic 2-point signal is overwhelmed on large-scales by the standard perturbations, with the $B$-mode polarisation being perhaps the most realistic option if we are to detect it directly. The increasing accuracy of measurements of the CMB non-Gaussianity provides an alternative. The stress tensor of a magnetic field is non-linear, implying that the statistics induced on matter perturbations are intrinsically non-Gaussian, regardless of the nature of the underlying magnetic field. Since the standard scenario contains relatively few sources of primordial non-Gaussianity, it is possible that a magnetic signal is dominant. Viewed another way, predicted signals from a magnetic field are likely to be of a characteristic nature, and must be found in and cleaned from the CMB data before any conclusions on early-universe physics can be made.

Aspects of the three-point moments have been studied in a series of papers in the last few years \citep{Brown:2005kr,Brown:2006wv,Seshadri:2009sy,Caprini:2009vk,Trivedi:2010gi,Cai:2010uw,Shiraishi:2010yk,Kahniashvili:2010us}. A bispectrum is set by three wavevectors, which we denote with $\mathbf{k}$, $\mathbf{p}$ and $\mathbf{q}$. Since to retain statistical isotropy these must form a closed triangle, this geometry can equivalently be expressed with the scalars $k,r,\phi$, where $p=rk$ and $\phi$ is the angle between $\mathbf{p}$ and $\mathbf{q}$. Employing these variables the bispectrum geometry can be written as a foliation of planes of constant $r$ and for each constant angle $\phi$ we then have a one-dimensional line through the bispectrum which in broad terms is expected to act in a similar manner to the power spectra.

The magnetic bispectra studied thus far have typically been along only three such lines, all in the $r=1$ plane -- the ``colinear'' case where $k=p=q/2$ and so $\phi=0$ \citep[hereafter BC05, B06 and CFPR09]{Brown:2005kr,Brown:2006wv,Caprini:2009vk}, the ``equilateral'' case where $k=p=q$ and so $\phi=2\pi/3$ \citep[hereafter SS09, CFPR09 and TSS10]{Seshadri:2009sy,Caprini:2009vk,Trivedi:2010gi}, and the ``local'' or degenerate case where $k\approx p$, $q\approx 0$ and so $\phi\approx\pi$ (SS09, CFPR09, TSS10). TSS10 also considered configurations where $\phi=0$ but $k\neq p$.

The recent studies have expanded the previous results considerably. SS09 considered the equilateral and degenerate lines of the bispectrum of the magnetic energy density for nearly scale-invariant magnetic fields, concluding that the degenerate line provides the greatest contribution to the integral and employing an approximation to this dominant term to estimate the CMB signal. Likewise, CFPR09 considered the bispectrum of the energy density and considered the colinear, equilateral and degenerate lines. The authors generally relied on approximations that neglect angular terms in the integrations, or apply only on large scales. Doing so recovers the scaling behaviour of the bispectrum at the expense of an accurate calculation of the relative amplitudes between lines. Since the degenerate line was found to diverge as $q^{2n+3}$ as $q\rightarrow 0$ this term is likely to dominate. CFPR09 also present exact solutions for the colinear case for both a causal field and a field relatively close to scale-invariance, which enable them to test their approximations. The approximations are certainly reasonable, but not ideal. In particular, since the bispectra are not positive-definite it is unclear whether there are strong cancellations to the degenerate line arising from other parts of the bispectrum. \citet{Cai:2010uw} employed the approximations of SS09 and CFPR09 and extended the treatment to full transfer functions. More recently, TSS10 considered the bispectrum of the anisotropic pressure of a near scale-invariant magnetic field. Unlike the previous papers they evaluated the bispectrum along the degenerate line in full, without neglecting any angular terms, finding it to be positive (in contrast with the approximate result, which would be negative).

However, there are four components of the magnetic stress tensor comprising six degrees of freedom -- the isotropic pressure/energy density, the anisotropic pressure, a vector component, and a transverse-traceless tensor component. Each of these are of roughly equal magnitude, as evidenced by the two-point moments \citep[for example]{Paoletti:2008ck,Brown:2010ms}. There are many correlations that can arise between these at the three-point level, and in principle all of them must be considered for the CMB. The rotationally-invariant combinations are listed in BC05 and B06, and numerical solutions for the colinear line are given, although for $n_B<-1$ these took the form of noisy statistical realisations. \citet{Shiraishi:2010sm} recently presented a formalism capable of considering the vector and tensor auto-correlation in full generality; our previous work considered only a fully-contracted form of the tensor auto-correlation and neglected the vector auto-correlation entirely. In \citet{Shiraishi:2010yk} this formalism was applied to the 3-point vector auto-correlation. In this paper the authors employed a technique to integrate the full bispectrum and, therefore, evaluate an unambiguous CMB signal.\footnote{The authors followed this with a detailed presentation in \citet{Shiraishi:2011fi} while this manuscript was under review.} A similar paper was presented by \citet{Kahniashvili:2010us}. These authors focused on the symmetries of the intrinsic bispectrum and on the appearance of anisotropic and off-diagonal terms but in principle considered the full bispectrum.

It is this consideration of the full bispectrum that is lacking for the other auto-correlations. This causes problems for causal fields and other fields far from scale-invariance in particular, where it is not necessarily to be expected that power is concentrated on the degenerate line. In this paper I address the full bispectrum across the range of $\{k,r,\phi\}$. For numerical study I consider a white noise field and a field near to scale-invariance, and evaluate the three auto-correlations $\av{\tau^3}$, $\av{\tau_S^3}$ and $\av{\tau_T^3}$ representing the bispectra of the magnetic energy density, anisotropic pressure, and gravitational wave source respectively. It is also possible to find exact solutions for the degenerate line of fields far from scale-invariance, and I present the large-scale limits of these. (Interested readers can find the full solutions using the techniques presented here, or else contact the author for details.) These solutions complement two particular solutions presented for the colinear line by CFPR09. Due the complexity of the integration volume, also discussed in that work, I focus otherwise on numerical techniques.

In section \ref{MagneticFields} I present a brief overview of the model and in section \ref{GeneralConsiderations} set up the formulae necessary to find the bispectra. Sections \ref{Bulk}, \ref{Colinear} and \ref{Squeezed} consider the general case, the colinear case and the squeezed case respectively and then I present my results in section \ref{Results}. Section \ref{Conclusions} provides a brief conclusion. The appendices contain some additional formulae and some analytical solutions for the degenerate bispectrum on large scales.

\section{Tangled magnetic fields}
\label{MagneticFields}
At the linear level a large-scale primordial magnetic field $b_a(\mathbf{k},\eta)$ decays as $a^2(\eta)$. We work with the scaled field $B_a(\mathbf{k})=a^2(\eta)b_a(\mathbf{k},\eta)$, which is constant over time \citep{Subramanian:1997gi,Jedamzik:1996wp}, and assume the electric fields to be negligible. Magnetic fields are damped on small scales, primarily from radiation viscosity, and we approximately model this effect with a time-dependent cut-off scale $k_c(\eta)$ above which they have no power, approximating a sharp damping tail. Note that there is then a time-dependence in the magnetic field $B_a(\mathbf{x})$ associated with the damping scale which in this section we write explicitly. The damping scale freezes at photon decoupling and $k_c(\eta_0)\approx k_c(\eta_\mathrm{rec})$.

Take as a toy model Gaussian random magnetic fields with the power spectrum
\be
  \langle B_a(\mathbf{k},k_c(\eta))B^*_b(\mathbf{k}',k_c(\eta))\rangle=\mathcal{P}_B(k)P_{ab}(\mathbf{k})H(k_c(\eta)-k)(2\pi)^3\delta(\mathbf{k-k}')
\ee
where
\be
\label{MagneticSpectrum}
  P_{ab}(\mathbf{k})=\delta_{ab}-\frac{k_ak_b}{k^2}
\ee
projects objects onto a plane orthogonal to the wavevector $\mathbf{k}$, $\mathcal{P}_B(k)$ is the magnetic power spectrum and in the interests of simplicity I have neglected an antisymmetric, helical component. $H(x)$ is the Heaviside function, which is defined to vanish for $x<0$ and to be unity for $x\geq 0$. The power spectrum is defined as
\be
\label{MagneticPowerSpectrum}
  \mathcal{P}_B(k)=A_Bk^{n_B}.
\ee
The analyticity of $\mathcal{P}_BP_{ab}$ requires $n_B\geq 2$, corresponding to magnetic fields generated by causal processes \citep{Durrer:2003ja}. Inflation can produce fields with $n_B\rightarrow -3$, corresponding to near scale-invariance. Some current constraints (generally based on the two-point moments) constrain the spectral index to $n_B<-0.12$ \cite{Paoletti:2010rx} at 95\% confidence and $n_B=-2.37^{+0.88}_{-0.73}$ at $1$-$\sigma$. The amplitude of the power spectrum $A_B$ can be normalised to observation either by the mean square field \citep[for example]{Finelli:2008xh,Paoletti:2008ck,Caprini:2009vk,Bonvin:2010nr} or more commonly by the field smoothed on a scale $\lambda$ (see for example \citet{Mack:2001gc}, CFPR09 and B10) but this choice is widely prevalent in the literature). In this paper we do not make contact with observation and leave $A_B$ unfixed. Note, though, that the normalisation requires $n_B\geq -3$ to keep the integration finite.

The magnetic field contributes to the Euler and Einstein equations through the stress-energy tensor $\tau^\mu_\nu(\mathbf{k})$. As the Poynting vector vanishes to first order and the magnetic energy density is equivalent to the isotropic pressure it is sufficient to consider the stress tensor $\tau^a_b(\mathbf{k})$,
\be
  \tau^a_b(\mathbf{k})=\tilde{\tau}^i_i(\mathbf{k})\delta^a_b-\tilde{\tau}^a_b(\mathbf{k}),
\qquad \mathrm{where}
\quad
 \tilde{\tau}_{ab}(\mathbf{k})=\int B_a(\mathbf{k}')B_b({\mathbf{k-k}'})\dcubedk'
\ee
is the self-convolution of the magnetic field. The stress tensor can be separated into the isotropic pressure (scalar trace), anisotropic pressure (traceless scalar), vector and transverse-traceless (TT) tensor components, which we denote by $\tau$, $\tau_S$, $\tau_a^V$ and $\tau_{ab}^T$ respectively. For more details see B06. The magnetic field also contributes through the Lorentz force and this has been studied in B06 and \citep{Paoletti:2008ck,Paoletti:2010rx}; however, here we focus on the statistics of the stresses themselves.

Magnetic fields have been separated into ``ultra-violet'' and ``infra-red'' cases by the behaviour of their stress power spectra; the power spectra for $n_B\geq -3/2$ are dominated by the damping scale and referred to as ``ultra-violet'' fields, while the spectra for $n_B<-3/2$ are independent of $k_c(\eta)$ across a wide range of scales and referred to as ``infra-red'' fields. It was noted in \citet{Brown:2010ms} (hereafter B10) that this has consequences for the ``coherence'' of the stresses. Since the damping scale is time-dependant, the statistics induced by a magnetic field on the matter perturbations are in principle also time-dependant. In that paper I identified a coherence scale $k_\mathrm{Coh}$ below which the stresses are ``decoherent'' and the standard techniques for evaluating CMB signals is suspect. The coherence scale is of the order of the damping scale for infra-red fields, but that for the ultra-violet fields is on much larger scales with $k_\mathrm{Coh}\sim k_c(\eta)/100$. On larger scales the stresses then tend towards white noise.

It is to be expected that this qualitative behaviour will hold also for the bispectra. There will be an ultra-violet r\'egime in which the bispectra are dominated by the damping scale and only tend towards white noise for $k\lesssim k_{\mathrm{Coh},\mathrm{3}}$ with a low $k_{\mathrm{Coh},\mathrm{3}}$, and an infra-red r\'egime where the bispectra will scale as power laws. However, it is interesting to note that the transition between these r\'egimes will occur not at $n_B=-3/2$ but instead at $n_B=-1$. This implies that magnetic fields with $n_B\in(-3/2,-1)$ act decoherently with respect to the two-point moments but coherently with respect to the three-point moments.  Extending this to higher-order moments, an $a$-point moment divides the ultra-violet and infra-red r\'egimes at $n_B=-3/a$ and all spectral indices in $n_B\in(-3/2,0)$ can be treated coherently if one considers a correlation of sufficiently high order.

\section{Intrinsic Bispectra}
\subsection{General Considerations}
\label{GeneralConsiderations}
\begin{center}
\begin{figure}
\begin{center}\includegraphics[width=0.4\textwidth]{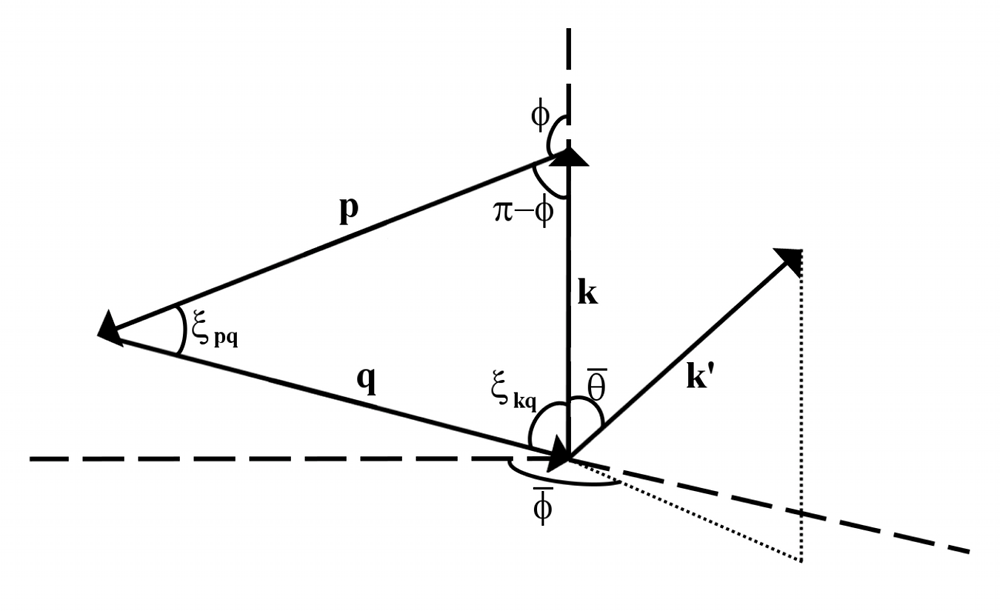}\end{center}
\caption{The geometry of a statistically homogeneous 3-point moment and the integration wavenumber $\mathbf{k}'.$}
\label{BispectrumGeometry}
\end{figure}
\end{center}

Since a bispectrum is constructed from three wavevectors $\mathbf{k}$, $\mathbf{p}$ and $\mathbf{q}$ forming a closed triangle, the bispectrum inhabits a three-dimensional space parameterised by three scalar quantities. I will typically take these to be $k=|\mathbf{k}|$, $r=|\mathbf{p}|/k$ and $\phi=\cos^{-1}\left(\khv\cdot\phv\right)$ (see Figure \ref{BispectrumGeometry}), although other parameterisations are possible (see for example \citet{Fergusson:2006pr} which employs a transformation to triangular coordinates, or SS09 and CFPR09 which employ $\{k,p,q\}$ directly). In this paper we will work in configurations where $r\geq 1$ and $\phi\in[0,\pi]$. This ensures that $k\leq p$ at all times, while in principle $q\in[0,\infty]$. Other configurations can be found by permutations of the wavevector labels. This is in contrast to SS09 and CFPR09 who explicitly make the permutations clear.

Bispectra are often evaluated in particular geometries where the three wavevectors form certain triangles, the most frequent of these being the equilateral case and the ``local'', ``degenerate'' or ``squeezed'' case in which $\mathbf{k}=-\mathbf{p}$ and $\mathbf{q}=0$. In BC05, B06, CFPR09 and TSS10 a third situation was considered, the ``colinear'' case where $\mathbf{k}=\mathbf{p}$ and so $\mathbf{q}=-2\mathbf{k}$. In terms of the coordinates $\left\{k,r,\phi\right\}$, these form lines of constant $\phi$ at $r=1$; the colinear line is at $\phi=0$, the equilateral line is at $\phi=2\pi/3$ and the degenerate line is at $\phi=\pi$. The other configurations in the $r=1$ plane describe isosceles configurations (as do certain other equivalent configurations when $\mathbf{p}=\mathbf{q}$).

Throughout this paper I use the following nomenclature:\newline
\begin{center}\begin{tabular}{r|l}
  ``Colinear line'' & $r=1$, $\phi=0$ \\
  ``Equilateral line'' & $r=1$, $\phi=2\pi/3$ \\
  ``Degenerate line'' & $r=1$, $\phi=\pi$ \\
  ``Colinear plane'' & $r>1$, $\phi=0$ \\
  ``Squeezed plane'' & $r>1$, $\phi=\pi$ \\
  ``Bulk'' & $r\in[1,\infty]$, $\phi\in(0,\pi)$
\end{tabular}\end{center}

The squeezed plane is often referred to as a ``local isosceles'' shape, while the degenerate line is referred to as a ``local'' or simply ``squeezed'' shape. The colinear line has also been referred to as  a ``midpoint collinear'' shape and the colinear plane as a ``squeezed collinear'' shape. See TSS10 for an example of such usage.

To calculate the CMB signal we wrap the intrinsic bispectrum across transfer functions; heuristically speaking, we have an integration of the form
\be
  B_{lmn}\sim \iiint f(k,p,q,l,m,n)\mathcal{B}(k,p,q)\Delta_{Tl}(k)\Delta_{Tm}(p)\Delta_{Tn}(q)k^2p^2q^2\dk\mathrm{d}p\dq
\ee
with $f(k,p,q,l,m,n)$ some function (which in general will include spherical Bessel functions and Wigner 3-j symbols). We therefore require in principle the full three-dimensional intrinsic bispectra, even though in SS09 and CFPR09 it was argued that the degenerate line dominates the integral for realistic situations. The calculations in these papers employed approximations neglecting parts of the integrals; as this can even cause a sign change the impact on the CMB signal could be significant. This was partially rectified in TSS10 where the degenerate line of the anisotropic scalar bispectrum was calculated exactly. However, the remaining parts of the bispectrum are still only known approximately, and the extent to which they modify the CMB signal is unknown. \citet{Shiraishi:2010yk} employed an alternative formalism and calculated the full integration for a vector auto-correlation.

The intrinsic magnetic bispectra of a Gaussian field can be written (see for example BC05) as
\bea
\label{GenericBispectra}
  \av{\tau_A(\mathbf{k})\tau_B(\mathbf{q})\tau_C(\mathbf{p})}&=&\delta(\mathbf{k}+\mathbf{p}+\mathbf{q})\int\Pow{k'}
    \Pow{\left|\mathbf{k}-\mathbf{k}'\right|}\Pow{\left|\mathbf{p}+\mathbf{k}'\right|} \nonumber \\ && \quad \times
    H(k_c(\eta)-k')H(k_c(\eta)-\left|\mathbf{k}-\mathbf{k}'\right|)H(k_c-\left|\mathbf{p}+\mathbf{k}'\right|)
    \left(8\mathcal{F}_{ABC}(\Theta)\right)\dcubedk'
\eea
where $\left\{ABC\right\}$ denote denote different parts of the stress-energy tensor, combined in rotationally-invariant combinations, and $\mathcal{F}_{ABC}(\Theta)$ is an angular component which is generally somewhat convoluted in form. $\Theta$ is the set of angle cosines which contains
\be
\begin{array}{c}
 \theta_{kp}=\khv\cdot\phv, \quad \theta_{kq}=\khv\cdot\qhv, \quad \theta_{pq}=\phv\cdot\qhv , \\
 \alpha_k=\khv\cdot\khv', 
 \quad \beta_k=\khv\cdot\widehat{\mathbf{k}-\mathbf{k}'}, 
 \quad \gamma_k=\khv\cdot\widehat{\mathbf{p}+\mathbf{k}'},
\end{array}
\ee
with equivalent expressions for $\alpha_p$, $\beta_p$, $\gamma_p$, $\alpha_q$, $\beta_q$ and $\gamma_q$, and
\be
  \betab=\khv'\cdot\mathbf{\widehat{k-k'}} \quad \gammab=\khv'\cdot\mathbf{\widehat{p+k'}}\quad \mub=\mathbf{\widehat{k-k'}}\cdot\mathbf{\widehat{p+k'}} .
\ee 

In terms of the angles $\xi_{kq}$ and $\xi_{pq}$ in Figure \ref{BispectrumGeometry}, $\theta_{kq}=-\cos\xi_{kq}$ and $\theta_{pq}=-\cos\xi_{kq}$, with $\alpha_k=\cos\thetab$. I focus in this paper on the $\langle\tau(k)\tau(p)\tau(q)\rangle$, $\langle\tau_S(k)\tau_S(p)\tau_S(q)\rangle$ and $\langle\tau^T_{ij}(k)\tau^{Tj}_k(p)\tau_T^{ik}(q)\rangle$ auto-correlations. The angular components for the scalar correlations are given in Appendix \ref{Appendix-AngularTerms}; that for the tensor mode is extremely complex and can be found in the appendix of B06. A full list of the angular components of the rotationally-invariant correlations and cross-correlations, including correlations with the vector pieces, can also be found there.\footnote{The correlations presented in BC05 were incomplete and lacked the unwieldy tensor correlations.} The formalism of \citet{Shiraishi:2010sm,Shiraishi:2010kd} is capable of dealing with the CMB bispectra of more general vector and tensor correlations.

To tackle this equation, first write the wavevectors in units of the damping scale $k_c$. Then if $\{k,r,\phi\}$ are the input parameters, $p=rk$ and $\thetab$ and $\phib$ are the polar angles of the coordinate system, the integration mode can be written $\mathbf{k}'=\mathbf{a}k_c(\eta)$, and
\be
\label{BStep1}
  \mathcal{B}=k_c^3\int_{a=0}^1a^2\Pow{a}H(1-a)\int_{\alpha_k=-1}^1\int_{\phib=0}^{2\pi}\Pow{\left|\mathbf{k-a}\right|}\Pow{\left|\mathbf{p+a}\right|}
  H(1-\left|\mathbf{k-a}\right|)H(1-\left|\mathbf{p+a}\right|)
  (8\mathcal{F})\dphib\dalpha_k\da
\ee
where $\mathbf{k}'=\mathbf{a}k_c$ and
\beas
 & \left|\mathbf{k-a}\right|^2=k^2+a^2-2ak\alpha_k, \quad
  \left|\mathbf{p+a}\right|^2=p^2+a^2+2ap\alpha_p, & \\
 & \alpha_k=\cos\left(\overline{\theta}\right), \quad \alpha_p=\alpha_k\cos(\phi)+\alphaks\sin(\phi)\cos\left(\phib\right), \quad \alphaks=\sqrt{1-\alpha_k^2} . &
\eeas
This equation can be approached following a method similar to that in B10. This is slightly complicated by the extra integration across $\phib$. Note, though, that $\phib$ appears only in combination with $\sin(\phi)$. The cases $\phi=0$, $\phi\in(0,\pi)$ and $\phi=\pi$ are then distinct from one-another and can be considered separately. This corresponds to a separation into the bulk, the colinear plane and line, and the squeezed plane and degenerate line.

\subsection{$\phi\in(0,\pi)$}
\label{Bulk}
The integral (\ref{BStep1}) is better cast as an integration across $\alpha_p$ instead of $\phib$. To do so it is useful to find an expression for $\sin\phib$. Writing $\cos\phib$ in terms of $\alpha_p$ gives $\sin\phib=\pm\sqrt{\alphaks^2\sin^2\phi-(\alpha_p-\alpha_k\cos\phi)^2}/\alphaks\sin\phi$. Then when $\phib\in[0,\pi)$, $\sin\phib>0$ and while $\phib\in[\pi,2\pi)$, $\sin\phib<0$.

Consider first the case where $\phib\in[0,\pi)$. Holding $\alpha_k$ constant, $d\alpha_p/d\phib=-\sqrt{\alphaks^2\sin^2\phi-(\alpha_p-\alpha_k\cos\phi)^2}$ and so
\be
  \mathcal{B}=\frac{k_c^3}{2p}\int_{a=0}^1a\Pow{a}\int_{\alpha_k=-1}^1\Pow{\left|\mathbf{k-a}\right|}\int_{\alpha_p=\alpha_{p\pi}}^{\alpha_{p0}}\Pow{\left|\mathbf{p+a}\right|}\frac{(8\mathcal{F})}{\sqrt{\Psi}}\dalpha_p\dalpha_k\da
\ee
where
\be
  \Psi=\alphaks^2\sin^2\phi-(\alpha_p-\alpha_k\cos\phi)^2
\ee
and
\be
  \alpha_{p\pi}=\alpha_k\cos\phi-\alphaks\sin\phi, \qquad
  \alpha_{p0}=\alpha_k\cos\phi+\alphaks\sin\phi .
\ee
Now set $z=\left|\mathbf{p+a}\right|^2$, transforming the integral to
\be
 \mathcal{B}=\frac{k_c^3}{2p}\int_{a=0}^1a\Pow{a}H(1-a)\int_{\alpha_k=-1}^1\Pow{\left|\mathbf{k-a}\right|}H(1-\left|\mathbf{k-a}\right|)\int_{z=z_\pi}^{z_0}\Pow{\sqrt{z}}H(1-z)\frac{(8\mathcal{F})}{\sqrt{\Psi}}\dz\dalpha_k\da .
\ee 
Here
\be
  z_\pi=p^2+a^2+2ap(\alpha_k\cos\phi-\alphaks\sin\phi), \qquad
  z_0=p^2+a^2+2ap(\alpha_k\cos\phi+\alphaks\sin\phi)=z_\pi+4ap\alphaks\sin\phi
\ee
and
\be
  \alpha_p=\frac{z-p^2-a^2}{2p} .
\ee

Considering the other case where $\phib\in[\pi,0)$, the derivative instead becomes $d\alpha_p/d\phib=\sqrt{\alphaks^2\sin^2\phi-(\alpha_p-\alpha_k\cos\phi)^2}$ . Following the same process, the resulting integral is the same as in the previous case and so in general
\be
 \mathcal{B}=\frac{k_c^3}{p}\int_{a=0}^1a\Pow{a}H(1-a)\int_{\alpha_k=-1}^1\Pow{\left|\mathbf{k-a}\right|}H(1-\left|\mathbf{k-a}\right|)\int_{z=z_\pi}^{z_0}\Pow{\sqrt{z}}H(1-z)\frac{(8\mathcal{F})}{\sqrt{\Psi}}\dz\dalpha_k\da .
\ee

Now, at $z=z_\pi$ and at $z=z_0$, $\Psi=0$. There are then poles at the limits of the integration. Between these limits, $\Psi>0$; transforming $z$ to $z=z_\pi+2apf$ gives $\alpha_p=\alpha_k\cos\phi-\alphaks\sin\phi+f$ and $\Psi=f(2\alphaks\sin\phi-f)$. $\Psi$ is therefore positive. The quantity $f$ reaches a maximum at $z=z_0$, where $f_\mathrm{max}=2\alphaks\sin\phi$. As a result, $\Psi>0$ except at $z\in\{z_\pi,z_0\}$ where $\Psi=0$.

Finally, the integral across $\alpha_k$ can be transformed into one across $\left|\mathbf{k-a}\right|^2$. Setting $y=\left|\mathbf{k-a}\right|^2$ rapidly produces
\be
\label{BulkBispectrum}
  \mathcal{B}=\frac{k_c^3}{2kp}\int_{a=0}^1\Pow{a}H(1-a)\int_{y=(k-a)^2}^{(k+a)^2}\Pow{\sqrt{y}}H(1-y)\int_{z=z_\pi}^{z_0}\Pow{\sqrt{z}}H(1-z){\sqrt{\Psi}}(8\mathcal{F})dzdyda .
\ee
With $q=\sqrt{k^2+p^2+2kp\cos\phi}=k\sqrt{1+r^2+2r\cos\phi}$, the angles that contribute to the integrations can be summarised as
\beas
&  \theta_{kp}=\cos\phi, \quad \theta_{pq}=-\dfrac{k}{q}\left(1+r\cos\phi\right), \quad \theta_{pq}=-\dfrac{k}{q}\left(r+k\cos\phi\right), &\\
&  \alpha_k=\dfrac{k^2+a^2-y}{2ak}, \quad \alpha_p=\dfrac{z-p^2-a^2}{2ap}, \quad \alpha_q=-\dfrac{k}{q}\left(\alpha_k+r\alpha_p\right), &\\
&  \beta_k=\dfrac{k-a\alpha_k}{\sqrt{y}}, \quad \beta_p=\dfrac{k\theta_{kp}-a\alpha_p}{\sqrt{y}}, \quad \beta_q=\dfrac{k\theta_{kq}-a\alpha_q}{\sqrt{y}}, &\\
&  \gamma_k=\dfrac{p\theta_{kp}+a\alpha_k}{\sqrt{z}}, \quad \gamma_p=\dfrac{p+a\alpha_p}{\sqrt{z}}, \quad \gamma_k=\dfrac{p\theta_{pq}+a\alpha_q}{\sqrt{z}}, &\\
&  \betab=\dfrac{k\alpha_k-a}{\sqrt{y}}, \quad \gammab=\dfrac{p\alpha_p+a}{\sqrt{z}}, \quad \mub=\dfrac{kp\theta_{kp}+a(k\alpha_k-p\alpha_p-a)}{\sqrt{yz}} . &
\eeas

\subsection{$\phi=0$}
\label{Colinear}
When $\phi=0$ the dependence on $\phib$ vanishes and the integral across $\phib$ becomes trivial. In this alignment $\alpha_k=\alpha_p$ and so
\be
  \left|\mathbf{k-a}\right|^2=k^2+a^2-2ak\alpha_k, \quad
  \left|\mathbf{p+a}\right|^2=p^2+a^2+2ap\alpha_k .
\ee
These become zero at $\alpha_k^0=(k^2+a^2)/2ak$ and $\alpha_p^0=-(p^2+a^2)/2ap$ respectively. Since these occur on either side of $\alpha_k=0$ the integration over $\alpha_k$ can be separated into two regions, transforming $\left|\mathbf{p+a}\right|$ when $\alpha_k<0$ and $\left|\mathbf{k-a}\right|$ when $\alpha_k>0$. Doing so yields
\bea
  \mathcal{B}&=&\pi k_c^3\int_0^1a\Pow{a}H(1-a)\left(
    \frac{1}{p}\int_{(p-a)^2}^{p^2+a^2}\Pow{\left|\mathbf{k-a}\right|}\Pow{\sqrt{y_p}}H(1-\left|\mathbf{k-a}\right|)H(1-y_p)(8\mathcal{F})\dy_p \right. \nonumber \\ && \left. \quad
    +\frac{1}{k}\int_{(k-a)^2}^{k^2+a^2}\Pow{\sqrt{y_k}}\Pow{\left|\mathbf{p+a}\right|}H(1-y_k)H(1-\left|\mathbf{p+a}\right|)(8\mathcal{F})\dy_k\right)\da
\label{ColinearSheetBispectrum}
\eea
where
\be
  \left|\mathbf{k-a}\right|^2=k^2+a^2+\frac{k}{p}\left(p^2+a^2-y_p\right), \quad
  \left|\mathbf{p+a}\right|^2=p^2+a^2+\frac{p}{k}\left(k^2+a^2-y_k\right) .
\ee
The angle cosines in the colinear limit reduce to
\be
\label{Angles-Colinear}
 \Theta_{kp}=-\Theta_{kq}=-\Theta_{pq}=1, \quad \alpha_k=\alpha_p=-\alpha_q, \quad \beta_k=\beta_p=-\beta_q, \quad \gamma_k=\gamma_p=-\gamma_q
\ee 
and the angular terms are presented in equations (\ref{TsTsTs-Colinear})-(\ref{TtTtTt-Colinear}). In particular, note that the tensor auto-correlation explicitly vanishes identically, as was argued before in BC05.

In the colinear limit where $r=1$, this simplifies further. Under the transformation $y_k\rightarrow y_p$ the angular terms $\FTrTrTr$, $\FTsTsTs$ and $\FTtTtTt$ are invariant even though the underlying angles change; therefore when $r=1$ and $\phi=0$ the bispectrum becomes
\be
\label{ColinearLineBispectrum}
  \mathcal{B}=\frac{2\pi k_c^3}{k}\int_{a=0}^1a\Pow{a}H(1-a)\int_{y=(k-a)^2}^{k^2+a^2}\Pow{\sqrt{y}}\Pow{\sqrt{2k^2+2a^2-y}}\left(8\mathcal{F}\right)H(1-y)H\left(1+y-2(k^2+a^2)\right)\dy\da .
\ee

\subsection{$\phi\rightarrow\pi$}
\label{Squeezed}
\begin{figure}
\begin{center}\includegraphics[width=0.6\textwidth]{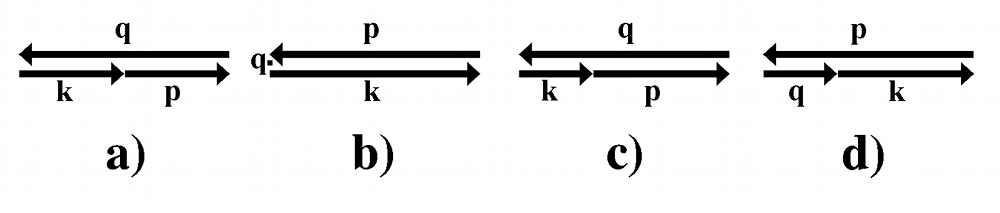}\end{center}
\caption{Bispectra at $\phi=0$ and $\phi=\pi$. a) The colinear configuration. b) The degenerate configuration. c) $\phi=0$. d) $\phi=\pi$. Note that case d) can be transferred into case c) if we map $\mathbf{q}\rightarrow\mathbf{k}$, $\mathbf{k}\rightarrow\mathbf{p}$, $\mathbf{p}\rightarrow\mathbf{q}$.}
\label{1DBispectra}
\end{figure}

There is a strong, intuitive difference between a squeezed bispectrum with $r=1$, which corresponds to the vanishing of $\mathbf{q}$, and a squeezed bispectrum with $r>1$. At $r=1$, the configuration is approaching a degenerate state where $q\rightarrow 0$ and at $\phi=\pi$ reaches $k=-p$, $q=0$. At $r>1$, however, the bispectrum is in a squeezed state with $q=p-k$. The symmetries of the bispectrum equate these squeezed configurations with an equivalent colinear bispectrum. Specifically, setting
\be
  \mathbf{k}\rightarrow\mathbf{p}, \quad \mathbf{p}\rightarrow\mathbf{q}, \quad \mathbf{q}\rightarrow\mathbf{k}
\ee
maps $\mathcal{B}(k,p/k,\pi)$ onto $\mathcal{B}(q,k/q,0)$. See Figure \ref{1DBispectra}.

As a result we consider the degenerate line separately from the rest of the squeezed plane.

\subsubsection{The Degenerate Line}
Here $r=1$ and $\phi=\pi$. In SS09 and CFPR09 it was shown that for $n_B<-1$ the bispectrum along this line is divergent as $q\rightarrow 0$ with $\mathcal{B}\propto q^{2n_B+3}$ to leading order, so we must restrict ourselves to $n_B\geq -1$. Since $\alpha_k=-\alpha_p$ one can write
\be
\label{DegenerateLineBispectrum}
 \mathcal{B}=\frac{k_c^3}{2k}\int_{a=0}^1a\Pow{a}H(1-a)\int_{(k-a)^2}^{(k+a)^2}\left(\Pow{\sqrt{y}}\right)^2H(1-y)\int_{\phib=0}^{2\pi}(8\mathcal{F})\dphib\dy\da .
\ee
The integration across $\phib$ remains since $\alpha_q=\alpha_q(\phib)$. Here $p=k$ and $q=0$. The vanishing of $q$ introduces spurious poles in the integrand which can be removed by Taylor-expanding around $\phi=\pi$, leading to
\bea
  \theta_{kp}=-1, & \theta_{kq}=0, & \theta_{pq}=0, \\
  \alpha_k=\cos\thetab, & \alpha_p=-\alpha_k, & \alpha_q=-\sin\thetab\cos\phib, \nonumber \\
  \beta_k=\dfrac{k-a\alpha_k}{\left|\mathbf{k-a}\right|}, & \beta_p=-\beta_k, & \beta_q=\dfrac{a\alphaks\cos\phib}{\left|\mathbf{k-a}\right|}, \\
  \gamma_k=-\beta_k, & \gamma_p=-\gamma_k, & \gamma_q=-\beta_q,  \nonumber \\
  \betab=\frac{k\alpha_k-a}{\left|\mathbf{k-a}\right|}, & \gammab=-\betab, & \mub=-1. \nonumber
\eea

The angular terms $\mathcal{F}$ are presented in equations (\ref{TrTrTr-Degenerate})-(\ref{TtTtTt-Degenerate}). The degenerate bispectrum in this case is similar to the power spectrum integrals evaluated in \citep{Finelli:2008xh,Paoletti:2008ck,Yamazaki:2008gr,Brown:2010ms} and should similarly possess exact solutions for certain values of $n_B$. Examination of equation (\ref{DegenerateLineBispectrum}) shows that the limits on the integration are the same as for the power spectrum employed in B10:
\begin{itemize}
  \item For $k\geq 2$, $k-a>0$ for all $a$ and so \be \mathcal{B}_{\mathrm{Deg}}(k\geq 2)\equiv 0. \label{DegLim}\ee
  \item For $k\in (1,2)$, $k+a>1$ for all $a$ and so $y_\mathrm{max}=1$. $k-a<1 \Rightarrow a>k-1$ and so
    \be \mathcal{B}_{\mathrm{Deg}}(k\in(1,2))=\frac{k_c^3}{2k}\int_{a=k-1}^1a\Pow{a}\int_{(k-a)^2}^1\left(\Pow{\sqrt{y}}\right)^2\int_{\phib=0}^{2\pi}\left(8\mathcal{F}\right)\dphib\dy\da \ee
  \item For $k\in(0,1]$, $(k+a)<1 \Rightarrow a<1-k$ and allows for $y_\mathrm{max}=(k+a)^2$, while $a\geq 1-k$ implies $y_\mathrm{max}=1$. Therefore \bea \mathcal{B}_{\mathrm{Deg}}(k\in(0,1])&=&\frac{k_c^3}{2k}\left(\int_{a=0}^{1-k}a\Pow{a}\int_{y=(k-a)^2}^{(k+a)^2}\left(\Pow{\sqrt{y}}\right)^2\int_{\phib=0}^{2\pi}(8\mathcal{F})\dphib\dy\da \right. \nonumber \\ && \qquad \left. +\int_{a=1-k}^{1}a\Pow{a}\int_{y=(k-a)^2}^{1}\left(\Pow{\sqrt{y}}\right)^2\int_{\phib=0}^{2\pi}(8\mathcal{F})\dphib\dy\da\right) \eea
\end{itemize}
Solutions exist for all integer and half-integer $n_B\geq -1/2$ and are presented for large scales in appendix \ref{AppendixDegenerate}; the full solutions are available on request or from the author's website\footnote{http://folk.uio.no/$\sim$ibrown}. These solutions complement the exact solutions for the colinear case of $\BTrTrTr$ presented in CFPR09, although they included a solution at $n_B=-2$ which due to the divergence with $q=0$ cannot be found for the degenerate line. Solutions for $n_B\lesssim -2$ are of particular interest as these fields are currently the weakest constrained \citep{Kahniashvili:2010wm,Paoletti:2010rx,Yamazaki:2010nf}.

If $n_B<-1$ this equation cannot be integrated; instead we sample finely along values of $\phi$ as $\phi\rightarrow\pi$ to track the divergence and set $\mathcal{B}(\phi=\pi)\equiv 0$ to impose numerical stability.

\subsubsection{The Squeezed Plane}
Since bispectra in this configuration can be mapped onto a colinear line we expect a very similar line. (We do not perform this mapping explicitly as our sampling in $\{k,r,\phi\}$ will not cleanly transfer to a reasonable sampling in $\{k',r',\phi'\}$. It is easier to integrate the squeezed plane directly than it is to refine the sampling to produce a suitably smooth result.) When $r>1$ and $\phi=\pi$, the dependence on $\phib$ drops out and so the bispectrum becomes
\be
  \mathcal{B}_{\mathrm{Sq}}=2\pi k_c^3\int_{a=0}^1a^2\mathcal{P}(a)H(1-a)\int_{\alpha_k=-1}^1\mathcal{P}(\left|\mathbf{k-a}\right|)\mathcal{P}(\left|\mathbf{p+a}\right|)H(1-\left|\mathbf{k-a}\right|)H(1-\left|\mathbf{k-a}\right|)(8\mathcal{F})\dalpha_k\da .
\ee
Consider the coordinate transformation $\left|\mathbf{k-a}\right|^2=y$. The modulus of $\mathbf{p+a}$ is then
\be
  \left|\mathbf{p+a}\right|^2=p^2+a^2+\frac{p}{k}(y_k-k^2-a^2)
\ee
which for fixed $r>1$ reaches zero at $k^2=p^2+ry_k$. This only occurs at $y_k=0$ which is at the edge of the integration volume; unlike the case for the colinear sheet there is no need to transform both moduli. The bispectrum is thus
\be
\label{SqueezedPlaneBispectrum}
  \mathcal{B}_{\mathrm{Sq}}=\frac{4\pi k_c^3}{k}\int_{a=0}^1a^2\mathcal{P}(a)H(1-a)\int_{y=(k-a)^2}^{(k+a)^2}\mathcal{P}(\sqrt{y})\mathcal{P}(\left|\mathbf{p+a}\right|)H(1-y)H(1-\left|\mathbf{p+a}\right|)(8\mathcal{F})\dalpha_k\da,
\ee
with
\be
  \left|\mathbf{p+a}\right|^2=p^2+a^2+\frac{p}{k}\left(y-k^2-a^2\right) .
\ee
As expected, this is similar to the colinear line.

\subsection{The Use of Statistical Realisations}
Following the techniques employed in BC05, B06 and B10 we can imagine finding the magnetic bispectra for both infra-red and ultra-violet fields by analysing realisations of cosmic magnetic fields. Equation (\ref{GenericBispectra}) was derived assuming that the magnetic fields obey an underlying Gaussian distribution; the use of realisations eases this restriction. In B06 we analysed the colinear configuration employing a grid of side-length $l_\mathrm{dim}=192$ and presented the results averaged from 1200 runs. These results took some time to recover and the errors rapidly grew dominant on large-scales; while we found good agreement between the realisations and numerical integration at $k\gtrsim 0.12k_c$ we excised the low-$k$ region. Since it is this very region that is significant for CMB studies, this is not an ideal solution.

Evaluating a complete bispectrum via realisations is certainly possible. However, as any general code must search through a grid for two complete wavemodes it is an extremely slow procedure. The scant mode-coverage on large scales forces one to employ the largest practical grid-size and number of realisations. Since bispectrum codes typically scale as $N^3\sim l_\mathrm{dim}^6$ this rapidly makes the use of realisations to generate a full 3D bispectrum unfeasible. Similar problems hit the attempt to employ realisations to study other one-dimensional linethroughs such as the degenerate or equilateral cases. In these situations the angular dependencies on $\phi$ force one to evaluate a subset of a full 3D bispectrum and the savings in time are not sufficient to render the problem readily tractable.

We can, however, test the analysis of the colinear line via realisations. In Figures \ref{LinethroughsN0} and \ref{LinethroughsNm25} the data points present the results from the average of 6000 realisations of a colinear magnetic bispectrum evaluated on grids of side-length $l_\mathrm{dim}=256$. Error bars are one standard deviation. For $n_B=0$ the results are in clear agreement with the numerical integration and a significant improvement over those previously presented. For $n_B=-5/2$ the bispectra, while vastly improved from those presented in BC05 and B06, exhibit a clear infra-red damping at $k<k_c$ arising from an unphysical infra-red cut-off. (This damping was discussed this damping in B10 and modelled semi-analytically with an IR-damped power spectrum.) For $k>k_c$ the realisations agree extremely well with the numerical integrations. Unfortunately the current technical restrictions make repeating this analysis for the rest of the bispectrum unrealistic.

\section{Results}
\label{Results}
The various bispectra, equations (\ref{BulkBispectrum}, \ref{ColinearLineBispectrum}, \ref{ColinearSheetBispectrum}, \ref{DegenerateLineBispectrum} and \ref{SqueezedPlaneBispectrum}), are integrated numerically employing a Miser Monte-Carlo routine\footnote{The Miser routine is adapted from that developed by Numerical Recipes {\tt www.nr.com}.}. I consider the cases $n_B=0$ and $n_B=-5/2$ as representative of ultra-violet and infra-red fields respectively.  Grid dependences were tested employing a range of numbers of samples in the chains; convergence was found to occur at approximately $n\sim 10^6$ for $n_B=0$ and $n\sim 10^7$ for $n_B=-5/2$. Results were found using $n=5\times 10^6$ for $n_B=0$ and $n=5\times 10^7$ for $n_B=-5/2$. The exception is the $\av{\tau_T^3}$ correlation where $n=5\times 10^8$ samples were needed for convergence across the entire plane due to the complexity of the integrand.

The grid the bispectra are integrated on also varies with $n_B$ but is relatively sparse in $k$, particularly for $n_B=-5/2$ where the most important area is $k\ll 1$ which is expected to exhibit a scaling behaviour. This sparse sampling is introduced to speed the numerics; for low $k$ the integration volume is large and integrations take quite some time, especially at low $n_B$. The angle $\phi$ is sampled once every five degrees except near the colinear and squeezed planes where it is sampled once every half a degree, and around the equilateral line for $\av{\tau_S^3}$ which for $n_B=-5/2$ contains a feature of particular interest. I consider $r\in[1,5]$.

Previous studies of the magnetic bispectra have typically been of scalar correlations, not least as it is only recently that a clear formalism for generating the CMB angular bispectra from vector or tensor perturbations has emerged \cite{Shiraishi:2010sm,Shiraishi:2010kd}. They have also been in $\{k,p,q\}$ space. Employing instead $\{k,r,\phi\}$ space has some benefits, the chief of which is the extent to which it highlights the scalings in $k$. In particular, the bispectra sampled along lines of constant $\{r,\phi\}$ are expected to obey a power-law in $k$ for $k\lesssim k_\mathrm{Coh}$. The bispectrum can then be modelled in a manner similar to the power spectra in, for example, B10 as a plane at constant $k$ and a given scaling relation. It also simplifies the construction of the complete bispectrum somewhat, in that we can integrate cleanly along lines of constant $\{r,\phi\}$ and assemble these lines into a 3D bispectrum. The other main benefit of employing this coordinate system is that it results in a fine sampling around the $r=1$ plane; from the results of SS09 and CFPR09 and the decaying of the bispectrum as $r\rightarrow\infty$, the degenerate line should dominate and other configurations in and close to the $r=1$ plane will be more significant than those further away.

To get the full bispectrum in $\{k,p,q\}$ space one can transform the results. Since they obey the triangle relations, bispectra in $\{k,p,q\}$ space are wedge-shaped. Transforming to this space yields a cone filling one third of the wedge. The other two thirds of the bispectrum in $\{k,p,q\}$ space can be found from permuting across $k$, $p$ and $q$.

There are, however, drawbacks to the coordinates $\{k,r,\phi\}$, the chief being a lack of immediate clarity. In particular, the $q^{2n_B+3}$ scaling found near to the degenerate line for $n_B<-1$ in CFPR09 is obscured.  Further, reconstructing the bispectrum in $\{k,p,q\}$ space will be slightly complicated when considering cross-correlations such as $\av{\tau(\mathbf{k})\tau(\mathbf{p})\tau_S(\mathbf{q})}$ since the anisotropy along the $\mathbf{q}$ vector must also be taken into account when performing the permutations. However, I feel that the convenience with which the bispectra can be found and, more importantly, the clean scalings in $k$ which should be recovered outweigh these disadvantages.

The bispectra are assembled with the following procedure:
\begin{itemize}
  \item For a set $\{r,\phi\}$ sample a bispectrum along a line in $k$.
  \item Repeat this for a set $r$ and samples in $\phi\in[0,\pi]$.
  \item Assemble these linethroughs into a plane at constant $r$, interpolating between missing points when necessary.
  \item Repeat this for samples in $r\in[1,5]$.
  \item Assemble the planes of constant $r$ into a 3D bispectrum.
\end{itemize}

There are a few general statements can be made about the form of a bispectrum before calculating the results. Along the colinear line the bispectrum will have power until $k=1$, which can be seen from evaluating the limits in equation (\ref{ColinearLineBispectrum}). Conversely, along the degenerate line, the bispectrum will have power until $k=2$, as seen from equation (\ref{DegLim}). We expect the power in the $r=1$ plane to smoothly connect the two. As a triangle on the squeezed plane can be rotated onto the colinear plane these parts of the bispectrum should be expected to resemble one-another, except when $r=1$ and the rotation breaks down.

In the same way, the limit $r\rightarrow\infty$ corresponds to the degenerate line on small scales since large $r$ implies that $k\ll p$ and $k\ll q$. The degenerate line itself will have a coherence length of $k_\mathrm{Coh}\lesssim 1$ and on smaller scales is decoherent. In the limit $r\rightarrow\infty$, then, the bispectrum should be expected to be decoherent across almost all scales. The coherence scale therefore obeys $k_\mathrm{Coh}(r\rightarrow\infty)\ll 1$. The conclusion is that the coherence length is a function of $r$ and decays for high $r$, although it is not necessarily a function of $\phi$.

Furthermore, since for near scale-invariant fields one expects $\mathcal{B}\sim k^{3n_B+3}$ across at least most of the bispectrum, the power on small scales is dramatically less than that on the largest scales. Another corollary of the equivalence of the high-$r$ r\'egime and the degenerate line is then that the power in the bispectrum should decay as $r\rightarrow\infty$. It will also damp entirely when $p>2$ which occurs at ever-smaller $k$ for increasing $r$. As $r$ increases, then, the bispectra for infra-red fields should lie within a narrowing wedge, decaying in amplitude at the same time. At higher $r$ the signal will be decoherent across almost the entire range. For such fields the decoherence at high $r$ is unlikely to be significant on the CMB.

For ultra-violet fields this is not the case since the large-scale behaviour is expected to be white noise, and the small and large scale signals are of approximately the same order of magnitude. While the coherence length will still evolve to larger scales as $r$ increases and the bispectrum damp away when $p>2$, the amplitude will remain of the same order-of-magnitude and the decoherence at high $r$ will be significant. This implies that our coordinates are not well-suited for the study of ultra-violet magnetic fields on the CMB. On the other hand, the caveats mentioned earlier and in B06 and B10 concerning the validity of wrapping decoherent statistics onto the CMB suggest we should take care with such cases regardless of the coordinate system employed.

The general structure of the bispectra should then be clear. There is power in the $r=1$ plane ranging from $k=k_c$ at the colinear line to $k=2k_c$ at the degenerate. For increasing $r$ the power dies at increasingly small scales until in the limit $r\rightarrow\infty$ it tends towards the small-scale (large-$k$) degenerate solution. The coherence length $k_\mathrm{Coh}$ likewise decays from a maximum in the $r=1$ plane until it vanishes at $r\rightarrow\infty$. The squeezed and colinear planes exhibit a similar behaviour to one-another although the quantitative nature is somewhat different. For infra-red fields the amplitude at high $r$ will be negligible, but for ultra-violet fields it will be comparable to that on the $r=1$ plane.

In analogy with the two-point moments of the stresses, on large scales a complete bispectrum for a typical magnetic power spectrum (such as the power-law considered here) will take the form
\be
\label{BispecScaling}
  \mathcal{B}(k,r,\phi)=\mathcal{B}_\star(r,\phi)\left(\frac{k}{k_\star}\right)^\alpha
\ee
for some pivot $k_\star\ll 1$. The values $\alpha=0$ for $n_B\geq-1$ and $\alpha=3n_B+3$ for $n_B<-1$ are expected, although this must be verified. This reduces the CMB integration to an integral across transfer functions and only a 2D plane rather than a 3D plane. In some situations it is possible that analytical approximations might be found given the power-law behaviour of the bispectrum. In any event, the full integration is then reduced to finding $\mathcal{B}_\star(r,\phi)$, the full knowledge of which is the factor missing in previous calculations. (Indeed, when TSS10 evaluated the degenerate line in full for an infra-red field they found that the full result has an opposite sign to the approximation neglecting $\mathcal{B}_\star(1,\pi)$.) It is far faster to integrate a plane with $\phi\in[0,\pi]$ and $r\in[1,r_\mathrm{max}]$ for constant $k$ than it is to integrate a full bispectrum. Such an increase in speed is vital if we are to perform accurate parameter estimation.

In principle the plane $\mathcal{B}_\star(r,\phi)$ should be tilted such that $k_\star(r)<k_\mathrm{Coh}(r)$, or chosen to be so small that $k_\mathrm{Coh}(r)$ only grows through $k_\star$ when the power in the bispectrum has become negligible. In practice, for infra-red fields the degenerate line is expected to dominate the bispectrum so significantly (SS09 and CFPR09, for example) that it is unlikely a significant error will be introduced by selecting a constant $k_\star\approx \mathcal{O}(10^{-4}-10^{-3})$.

In the results that follow, the errors presented are $1$-$\sigma$. These errors are dependant on the convergence of the integral as well as the number of samples taken in the average, owing to the Monte-Carlo nature of the integration.

\subsection{The Flat Spectrum, $n_B=0$}
Consider first the stresses of a white-noise magnetic field. This case is straightforward to integrate as it contains few of the poles that litter the integrations for infra-red fields. Analytical solutions are, however, extremely difficult to find and it is quite possible that they don't exist, with the exception of solutions along the degenerate line (see Appendix \ref{AppendixDegenerate}) although some solutions can be found in the small-scale tail. I integrate the equations numerically, setting a grid-spacing in $\phi$ of five degrees and employing $5\times 10^6$ samples in the Monte-Carlo chains. It is to be expected that the bispectra tend towards white noise ($\alpha=0$) for $k\ll 1$ with a coherence length on the order of $k_\mathrm{Coh}\approx 10^{-3}-10^{-2}$ in the $r=1$ plane. The magnitudes of the bispectra should be of the order of $(1-10)A_B$, although it is impossible to predict what configuration, if any, dominates.

\subsubsection{$\BTrTrTr$}
The auto-correlation of the isotropic pressure is straightforward. It is positive definite, showing a clean evolution between $\phi=0$ and $\phi=\pi$ and between $r=1$ and $r=5$. Figure \ref{LinethroughsN0} shows the colinear, equilateral and degenerate configurations. The solution for the colinear line agrees well with the results from realisations which are, unfortunately, extremely noisy at low $k$, albeit a great improvement over those previously presented.

\begin{figure}
\begin{center}\includegraphics[width=0.55\textwidth]{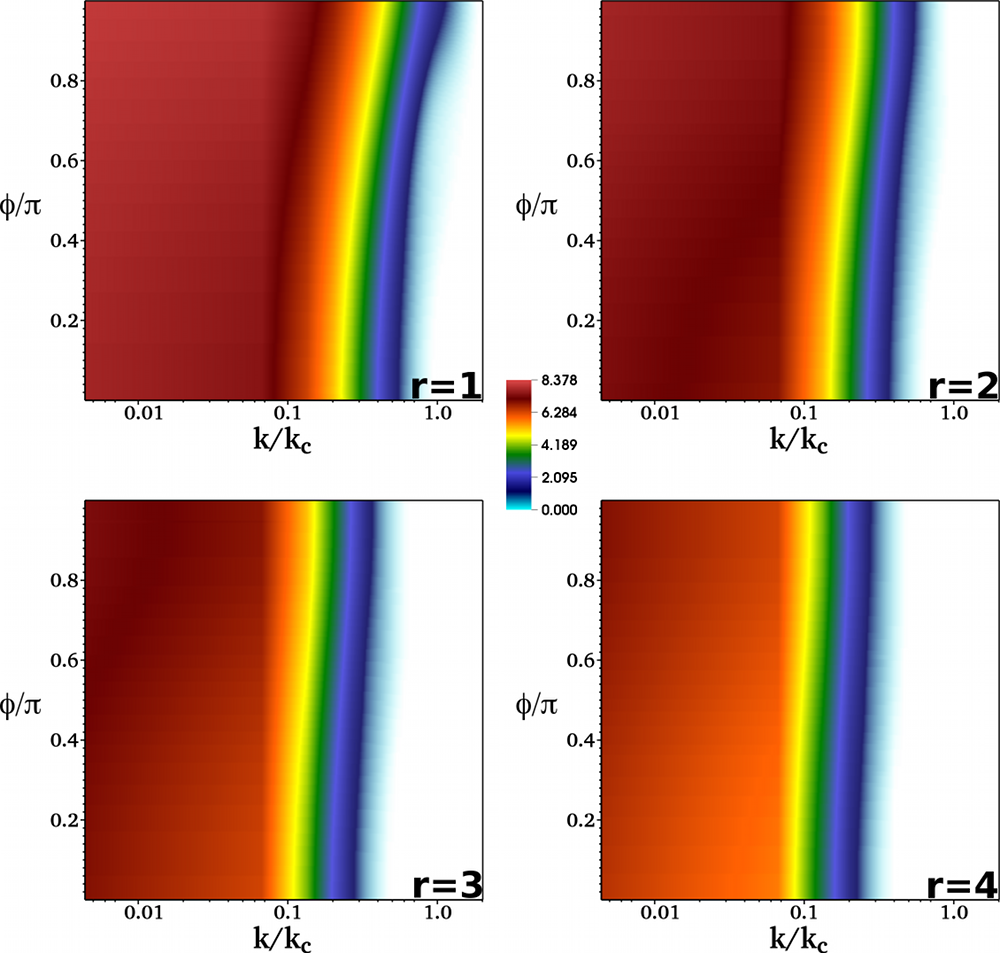}\end{center}
\caption{Slices through $\langle\tau^3\rangle$ for $n_B=0$ at constant $r$}
\label{PlanesN0TrTrTr}
\end{figure}

Figure \ref{PlanesN0TrTrTr} presents slices through the bispectrum at $r\in\{1,2,3,4\}$. In the plot for $r=1$ the transition between the colinear and degenerate behaviours is clear, with power decaying on ever-larger scales with increasing $\phi$ showing only a minor evolution for increasing $r$. The simplicity of $\mathcal{F}_{\tau\tau\tau}$ implies that integration is effectively constrained purely by the Heaviside functions in the power spectrum and not by cancellations in the integrands. The volume this bispectrum fills can then be taken as approximately marking the extremal bounds of any of the bispectra -- that is, power should not be expected outside of the bounds of $\av{\tau^3}$. Furthermore, the amplitude of the white-noise signal is almost constant with both $r$ and $\phi$. This is in agreement with and extends the results of CFPR09 who found that, for $n_B>-1$, the white noise amplitude is constant across the colinear, equilateral and squeezed configurations.

Figure \ref{PivotPlanesN0} gives the $(r,\phi)$ plane at a pivot scale $k_\star=7.5\times 10^{-5}$, which shows this strikingly. Modelling the bispectrum with a power-law around this plane, and for presentation choosing to evaluate $\alpha$ at the equilateral line, the integrations here give
\be
  \alpha=(0.16\pm 3.23)\times 10^{-4}
\ee
This is the result of sampling the average of six points in the white-noise r\'egime. Considering the full plane $\alpha(r,\phi)$ at $k=k_\star$ and low $r$ gives results which are similar. The coherence length evaluated at the equilateral line $k_\mathrm{Coh}$ evolves slowly towards larger scales from $k_\mathrm{Coh}(r=1)\approx 5\times 10^{-3}$ as $r$ increases. A r\'egime with $k_\mathrm{Coh}(r)\lesssim k_\star$ is reached at larger $r$. The recovered $\alpha$ would grow inaccurate unless $k_\star=k_\star(r)$, reducing at large $r$. The constant back plane and the white-noise behaviour of the scaling make this bispectrum particularly simple on large scales.

Approximating $\mathcal{B}_\star(r,\phi)=8.35$ and $\alpha=0$, then, with a pivot $k_\star(r)$ that always remains within $k_\mathrm{Coh}(r)$ produces a good model of the bispectrum as it is applicable to CMB scales. However, the rapid decay of $k_\mathrm{Coh}$ with $r$, combined with the relative strength of power on small scales compared to large scales, implies that one should be wary of wrapping these fields directly onto the CMB.

\begin{figure}
\begin{center}\includegraphics[height=0.9\textheight]{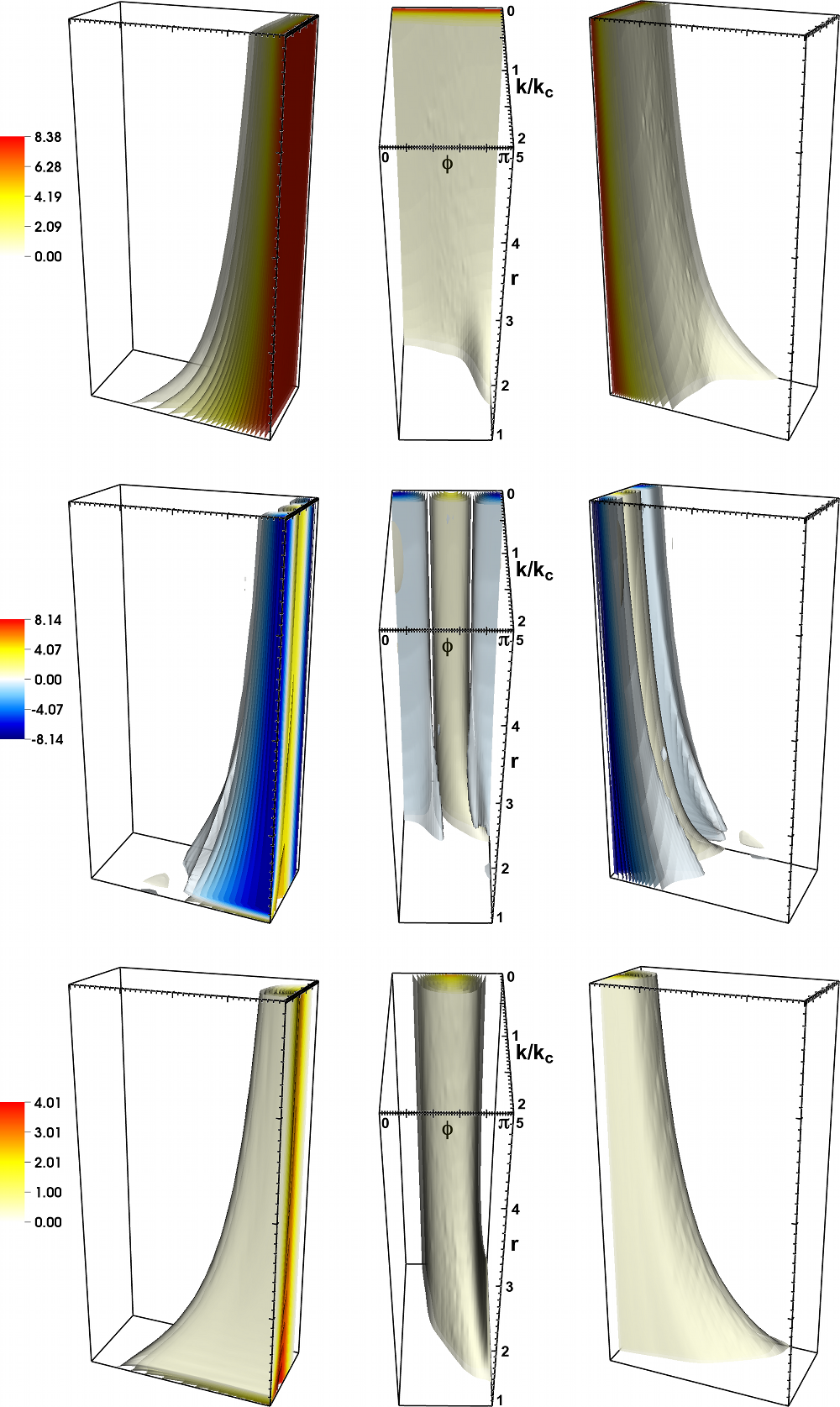}\end{center}
\caption{Isosurfaces through the auto-correlations $\tau^3$, $\tau_S^3$ and $\tau_T^3$ for $r\in[1,5]$ and $n_B=0$.}
\label{3DOut}
\end{figure}

Finally the full bispectrum in $k,r,\phi$ space is presented in Figure \ref{3DOut}. This plot shows isosurfaces of constant amplitude through the bispectrum. This figure highlights the simple sheet-like structure of this correlation, the wedge-shape of the bounds of the correlations and the relatively slow decay with increasing $r$.

\subsubsection{$\BTsTsTs$}
\begin{figure}
\begin{center}\includegraphics[width=0.98\textwidth]{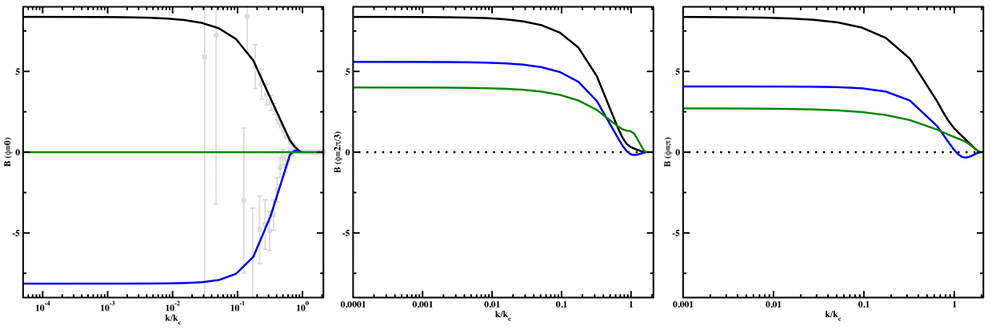}\end{center}
\caption{The colinear ($\phi=0$, left), equilateral ($\phi=2\pi/3$, middle) and degenerate ($\phi=\pi$) limits of the $\av{\tau^3}$ (black), $\av{\tau_S^3}$ (blue) and $\av{\tau_T^3}$ (green) bispectra, for $n_B=0$.}
\label{LinethroughsN0}
\end{figure}

\begin{figure}
\begin{center}\includegraphics[width=0.55\textwidth]{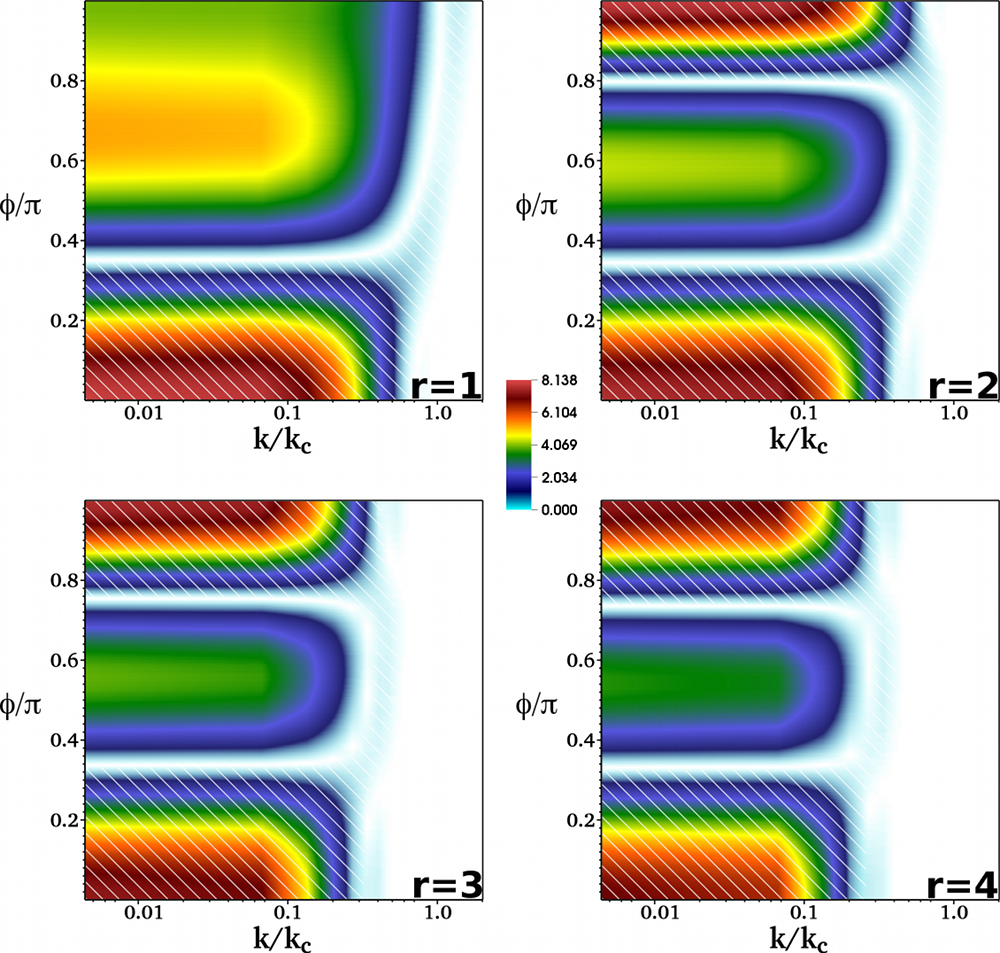}\end{center}
\caption{Slices through $\langle\tau_S^3\rangle$ for $n_B=0$ at constant $r$. Dashed areas are negative.}
\label{PlanesN0TsTsTs}
\end{figure}

The auto-correlation of the anisotropic is more interesting in structure than that of the isotropic pressure. In amplitude it is similar, and on white-noise scales is $A_{\tau_S}\approx 0.97A_\tau$, but it has a much more complicated pattern. Figure \ref{LinethroughsN0} shows that the signal is negative along the entire colinear line, while it is positive for much of the range of $k$ along the equilateral and degenerate lines. Even here, though, it is negative on small scales. The $\av{\tau_S^3}$ signal therefore passes between regions of positive and negative power dependant on both $\phi$ and $k$. It should be expected that the same is true for varying $r$ and this is confirmed in Figure \ref{PlanesN0TsTsTs}, which shows the negative region beside the colinear line, which is joined by a corresponding region along the squeezed plane. As expected, away from the degenerate line the squeezed plane has the same general behaviour as the colinear plane. The planes in Figure \ref{PlanesN0TsTsTs} slowly squeeze a positive region which is centred approximately around $\phi=\pi/2$. The coherence length at the equilateral line is $k_\mathrm{Coh}\approx 10^{-3}$, again evolving only very slowly with increasing $r$.

From the general arguments above, in the limit of high $r$ we would expect the bispectrum to tend to the large-scale degenerate solution. In Figure \ref{LinethroughsN0} we can see that this is negative, albeit extremely small. The central, positive, region is then expected to entirely vanish at the extreme reach of the bispectrum. As with the isotropic pressure, the signal along the colinear line is well-modelled with statistical realisations, but is extremely noisy on large scales.

Figure \ref{PivotPlanesN0} shows $\mathcal{B}_\star(r,\phi)$ at $k_\star=7.5\times 10^{-5}$ and demonstrates these features clearly. The region of positivity is asymmetric around the equilateral line at $\phi=2\pi/3$ with a tail at $r=1$ extending to reach the degenerate line at $\phi=\pi$. The bispectrum in general exhibits a characteristic ``triple drainpipe'' pattern. Recovering the scaling relation around the equilateral line gives
\be
  \alpha=(-1.73\pm 4.97)\times 10^{-4} .
\ee
The bispectrum can then be modelled by using $\mathcal{B}_\star(r,\phi)$ and $\alpha=0$; unfortunately this plane is not immediately amenable to analytic approximation. The full bispectrum in Figure \ref{3DOut} shows the triple-drainpipe extremely clearly, along with its slow decay at high $r$.

\subsubsection{$\BTtTtTt$}
As with the isotropic pressure, the bispectrum of the tensor anisotropic stress is positive-definite, with a peak magnitude of $A_{\tau_T}\approx 0.48A_{\tau}$. However, unlike the isotropic pressure, this signal has a non-trivial structure. We commented in BC05 that one can quickly prove that the colinear line of this signal is identically zero by the symmetries of the projection tensor, a conclusion which the full numerical integration confirms in Figure \ref{LinethroughsN0}, which shows the vanishing colinear signal, but the equilateral and degenerate lines being of the same order-of-magnitude as the scalar auto-correlations.

\begin{figure}
\begin{center}\includegraphics[width=0.55\textwidth]{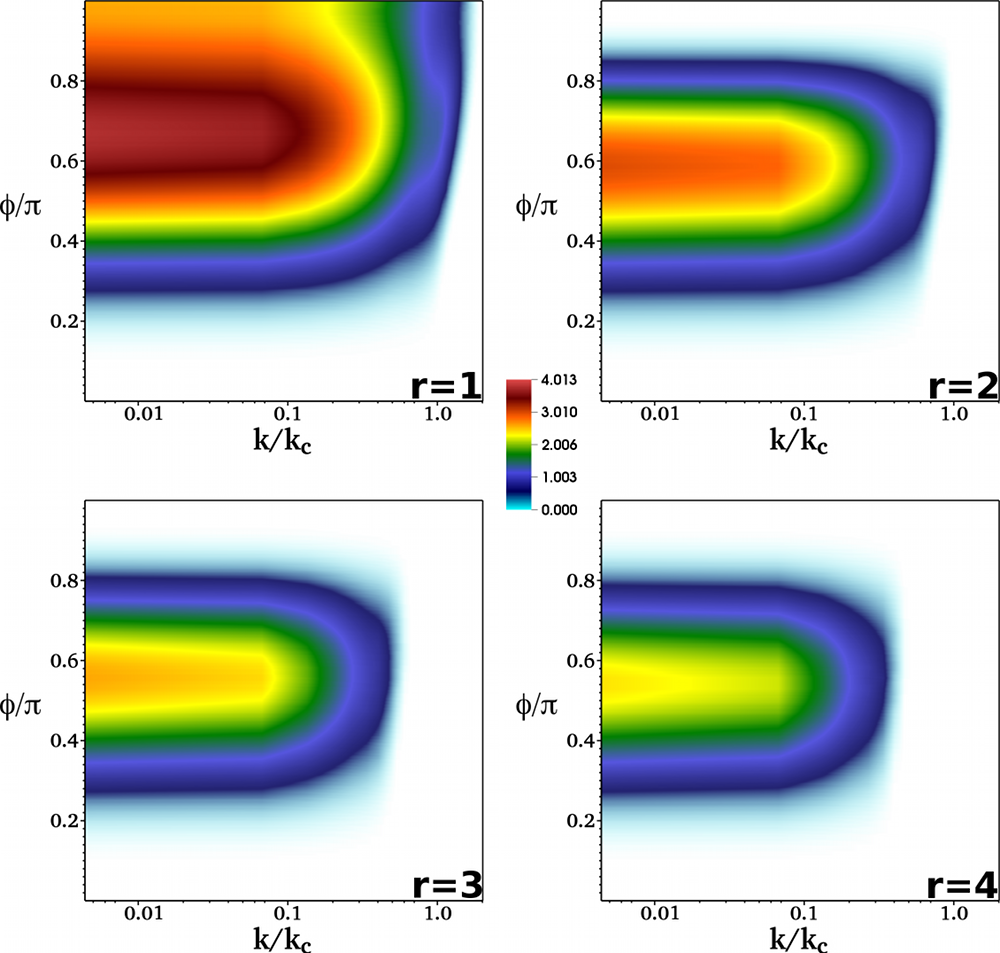}\end{center}
\caption{Slices through $\langle\tau_T^3\rangle$ for $n_B=0$ at constant $r$.}
\label{PlanesN0TtTtTt}
\end{figure}

Indeed, examining the plane at $r=1$ in Figure \ref{PlanesN0TtTtTt} reveals that the result is negligible (although, in principle, non-vanishing) up until $\phi\approx\pi/5$. For higher $r$ the symmetries between the squeezed and colinear planes imply that the squeezed plane should also vanish, and this is observed. By $r=4$ the signal vanishes for $\phi\lesssim\pi/4$ and $\phi\gtrsim 4\pi/5$. The maximum wavenumber at which a signal is non-vanishing is also strongly dependent on $\phi$, reaching its peak at $\phi\approx 0.57\pi$ for low $r$, diminishing slowly with increasing $r$. This selects a configuration of particular interest -- an isosceles lying between an equilateral and a right-angled triangle -- for this signal which could not easily be predicted from examining the angular integrand.

The coherence length at the equilateral line is $k_\mathrm{Coh}\approx 5\times 10^{-3}$ for $r=1$ and as with the scalar modes evolves only slowly.

\begin{figure}
\begin{center}\includegraphics[width=0.98\textwidth]{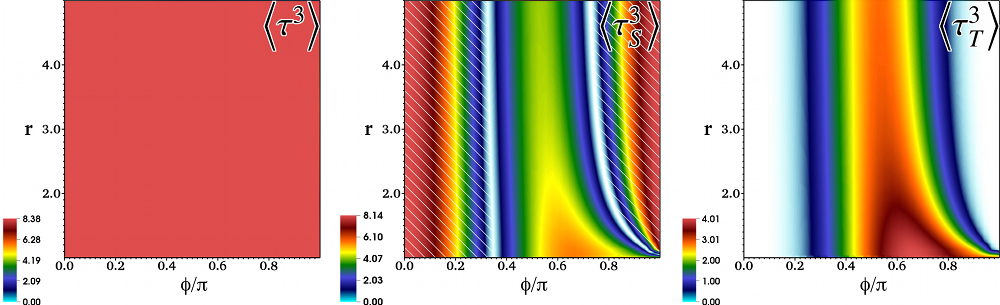}\end{center}
\caption{Slices through $\av{\tau^3}$, $\av{\tau_S^3}$ and $\av{\tau_T^3}$ at a pivot scale of $k_\star=4.4\times 10^{-3}k_c(\eta)$ and with $n_B=0$. The scale on $\av{\tau^3}$ has been truncated to show structure and dashed areas are negative.}
\label{PivotPlanesN0}
\end{figure}

The full bispectrum shown in Figure \ref{3DOut} reveals that, as $r$ increases, the region with a significant signal continues to narrow. It also highlights the characteristic shape of this signal, similar to a tree trunk, with a bulge on the base stretching out to the degenerate line. $\BTtTtTt$, plotted in Figure \ref{PivotPlanesN0}, shows that the power on large scales in this signal, in contrast to the two scalar auto-correlations, is focused in the vicinity of the $r=1$ plane and near-equilateral configurations. The high-$r$ solution is, however, similar in order-of-magnitude to the degenerate line, as should be expected. The scaling recovered around the equilateral line is
\be
  \alpha=(-1.52\pm 5.17)\times 10^{-4}.
\ee
The bispectrum can then be modelled with the plane $\mathcal{B}_\star(r,\phi)$ from Figure \ref{PivotPlanesN0} and $\alpha=0$.

\subsection{The Tilted Spectrum, $n_B=-5/2$}
Consider now the stresses of the ultra-violet field with $n_B=-5/2$. These integrals are much more difficult to integrate since the integration volumes are steeply tilted towards poles at the edges of the integration. While such integrals can be controlled numerically, much longer chains are required if convergence is to be reached: $5\times 10^7$ samples for the scalar modes, and $5\times 10^8$ for the tensor auto-correlation, which possesses a significantly more complicated integration surface. Analytical solutions are difficult to find and may well not exist. As with the white-noise field, $\phi$ is sampled once every five degrees, except around the equilateral line for $\av{\tau_S^3}$ which contains an interesting feature and is sampled at every half a degree.

The bispectra are expected to scale along lines of constant $\{\phi,r\}$ as $\mathcal{B}\propto k^{3(n_B+1)}$. This scaling has been observed for colinear (BC05, B06, CFPR09) and equilateral (CFPR09) bispectra and is na\"ively expected to hold throughout the rest of the bulk. Towards the degenerate line, SS09 and CFPR09 found divergences $\mathcal{B}\propto q^{2n_B+3}$ and concluded that this line dominates. In the $\{k,r,\phi\}$ coordinates this scaling will be obscured. Specifically, the dominant term found in CFPR09 was
\be
  \mathcal{B}\approx\delta(\mathbf{k+p+q})\frac{A_B^3}{144\pi^2}\left(\frac{2n}{(n_B+3)(2n_B+3)}k^nq^{2n_B+3}+\ldots\right);
\ee
after transformation this becomes
\be
  \mathcal{B}\approx\delta(\mathbf{k+p+q})\frac{A_B^3}{144\pi^2}\left(\frac{2n}{(n_B+3)(2n_B+3)}\left(1+r^2+2r\cos\phi\right)^{(2n_B+3)/2}k^{3(n_B+1)}+\ldots\right) .
\ee
The scaling in $k$ would then be expected to be the same as for the other configurations; however, a large drop in amplitude is expected with only a slight change in $r$ and $\phi$. For the field with $n_B=-5/2$ the scaling should be $\alpha=3(n_B+1)=-9/2$.

The stress power spectra on small scales for infra-red fields, $k\gtrsim 1$, are of the same order of magnitude as those for ultra-violet fields (see, for example, B10). This behaviour is expected to hold for the bispectra, too. Given the scaling on scales $k\lesssim k_\mathrm{Coh}$, very large signals in the r\'egime applicable to CMB studies are then expected.

In principle, the bispectra should lie within the same volume occupied by $\BTrTrTr(n_B=0)$. The strong scalings, and the expected domination of the degenerate line, will most likely obscure this feature.

\subsubsection{$\BTrTrTr$}
The 1D bispectra along the colinear, equilateral and a near-degenerate line are shown in Figure \ref{LinethroughsNm25} compared, in the colinear case, with the results from realisations. The agreement with the realisations is reasonable but the consequences of the infra-red cutoff are evident with the evaluation losing power on increasingly large scales. The signal is positive-definite, which agrees with the signs found in CFPR09 and SS09. Interestingly, the equilateral line is of the same order-of-magnitude as the colinear -- specifically, $\mathcal{B}_\mathrm{eq}/\mathcal{B}_\mathrm{col}\approx 1.02$. This is in conflict with the results of CFPR09 where the authors found that $\mathcal{B}_\mathrm{eq}/\mathcal{B}_\mathrm{col}\approx 0.1$. The discrepancy is almost certainly due to our inclusion of angular terms necessarily neglected in the approximations of CFPR09. In contrast, the bispectrum evaluated along a line at $\phi=35\pi/36$, corresponding to $\phi=175^\circ$, is three orders of magnitude greater. Lines nearer degeneracy are, naturally, larger again. As expected, the degenerate line will dominate the integration.

\begin{figure}
\begin{center}\includegraphics[width=0.55\textwidth]{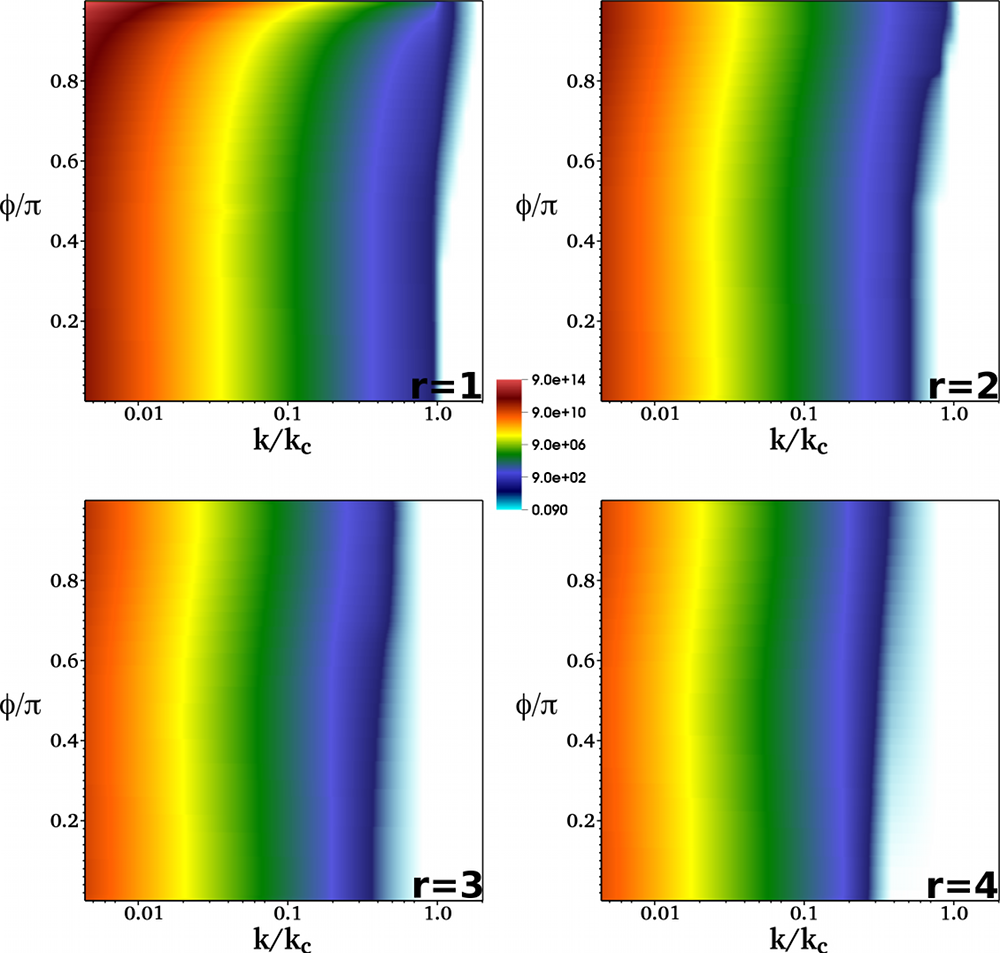}\end{center}
\caption{Slices through $\langle\tau^3\rangle$ for $n_B=-5/2$ at constant $r$.}
\label{PlanesNm25TrTrTr}
\end{figure}

Figure \ref{PlanesNm25TrTrTr} shows the bispectrum sliced at $r\in\{1,2,3,4\}$. These plots show the evolution of the bispectra from $\phi=0$ to $\phi=\pi$ clearly and the slowly-diminishing power as $r$ increases. The most dramatic feature is the area near to the degenerate line where the power diverges. By $r=2$ the squeezed plane has approximately equal power to the bulk. The lower curves in Figure \ref{LinethroughsNm25}, damped at small $k$, highlight this; these show the colinear, ``equilateral'' and squeezed configurations for $r=5$. While the colinear and ``equilateral'' signals are only approximately an order-of-magnitude lower than that at $r=1$, the squeezed signal is suppressed by five orders of magnitude.

The coherence length $k_\mathrm{Coh}$ evolves from $k_\mathrm{Coh}\approx 1$ at $r=1$ to $k_\mathrm{Coh}\approx 1/10$ at $r=5$ and therefore generally remains on much smaller scales than the CMB.

\begin{figure}
\begin{center}\includegraphics[width=0.98\textwidth]{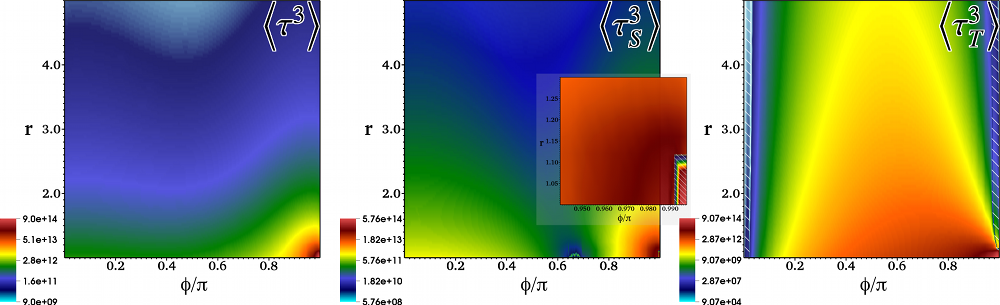}\end{center}
\caption{Slices through $\av{\tau^3}$, $\av{\tau_S^3}$ and $\av{\tau_T^3}$ at a pivot scale of $k_\star=4.4\times 10^{-3}k_c(\eta)$ and with $n_B=-5/2$. The scales have been truncated to highlight the structure, dashed areas are negative and the inset focuses on the vicinity of the degenerate line for $\av{\tau_S^3}$.}
\label{PivotPlanesNm25}
\end{figure}

Figure \ref{PivotPlanesNm25} shows $\mathcal{B}_\star(r,\phi)$ at a pivot scale of $4.4\times 10^{-3}$, where the pivot is taken to be larger pivot than for $n_B=0$ since $k_\mathrm{Coh}$ is reached at a smaller $k$. The scale has been truncated to show the increased structure in the pivot plane compared to the $n_B=0$ case. As with those $\av{\tau_S^3}$ and $\av{\tau_T^3}$ cases, this is unfortunately not immediately amenable to analytic approximation. Modelling the bispectrum with the power law around this plane we recover
\be
  \alpha=-4.498\pm 0.006
\ee
which is extremely close to the predicted $\alpha=-9/2$. Using the $\mathcal{B}_\star(r,\phi)$ plane with this value of $\alpha$ can then recover the section of the bispectrum relevant to the CMB.

The extent to which the power piles onto the degenerate line renders diagrams of the full bispectrum at $n_B=-5/2$ somewhat unhelpful. The most useful information can be extracted from the lane presented in Figure \ref{PivotPlanesNm25} and the recovered value of $\alpha$. It can also be noted that the increase of the error in $\alpha$ can be used to diagnose when $k_\mathrm{Coh}(r)$ has grown through $k_\star(r)$; however, given the domination of the degenerate line it is not expected that much error will be introduced at high $r$ from employing a constant $k_\star$.

\subsubsection{$\BTsTsTs$}
As with the previous bispectra (both for $n_B=0$ and $n_B=-5/2$), the $r=1$ plane can contain features that only exist right on this plane due to the high symmetries that can occur when $k=p$, the clearest examples being that the degenerate line for $n_B=0$ is positive for $\av{\tau_S^3}$ and non-vanishing for $\av{\tau_T^3}$, in contrast with the behaviour in the rest of the squeezed plane. The $\av{\tau_S^3}$ bispectrum at $n_B=-5/2$ illustrates this even more dramatically, although it also contains other unique features.

\begin{figure}
\begin{center}\includegraphics[width=0.98\textwidth]{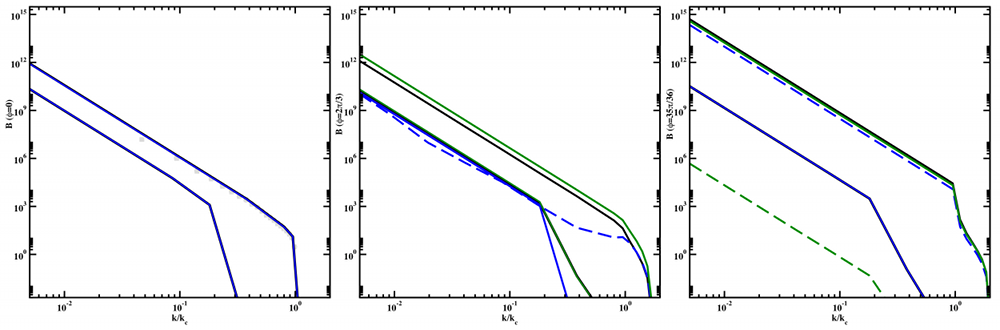}\end{center}
\caption{The colinear ($\phi=0$, left), equilateral ($\phi=2\pi/3$, middle) and near-degenerate ($\phi=0.977\pi$) limits of the $\av{\tau^3}$ (black), $\av{\tau_S^3}$ (blue) and $\av{\tau_T^3}$ (green) bispectra, for $n_B=-5/2$. Lines decaying for $k\gtrsim 1$ are at $r=1$ while those decaying at smaller $k$ are at $r=5$. Dashed lines are negative.}
\label{LinethroughsNm25}
\end{figure}

Figure \ref{LinethroughsNm25} shows the colinear, equilateral and near-degenerate lines. As with $\av{\tau^3}$ the agreement with realisations for the colinear line is good, with the realisations exhibiting an infra-red damping. The two scalar signals are indistinguishable from one-another along this line. Along the degenerate line, the signal has become negative, and almost equal in magnitude to $\av{\tau^3}$. This is in agreement with TSS10, recalling that their $\Pi_B=-\tau_S$, who found that their degenerate line was positive in contrast to the approximate solutions.

The equilateral line is more interesting. The signal along this line is significantly suppressed across much of the range of $k$ before flattening out as $k\rightarrow 1$. For $k>1$ it is of the same order of magnitude as the other two correlations. The bispectrum of the anisotropic pressure then exhibits some interesting behaviour in the $r=1$ plane, passing from positive to a negative region near the equilateral line. It remains negative near the degenerate line.

\begin{figure}
\begin{center}\includegraphics[width=0.55\textwidth]{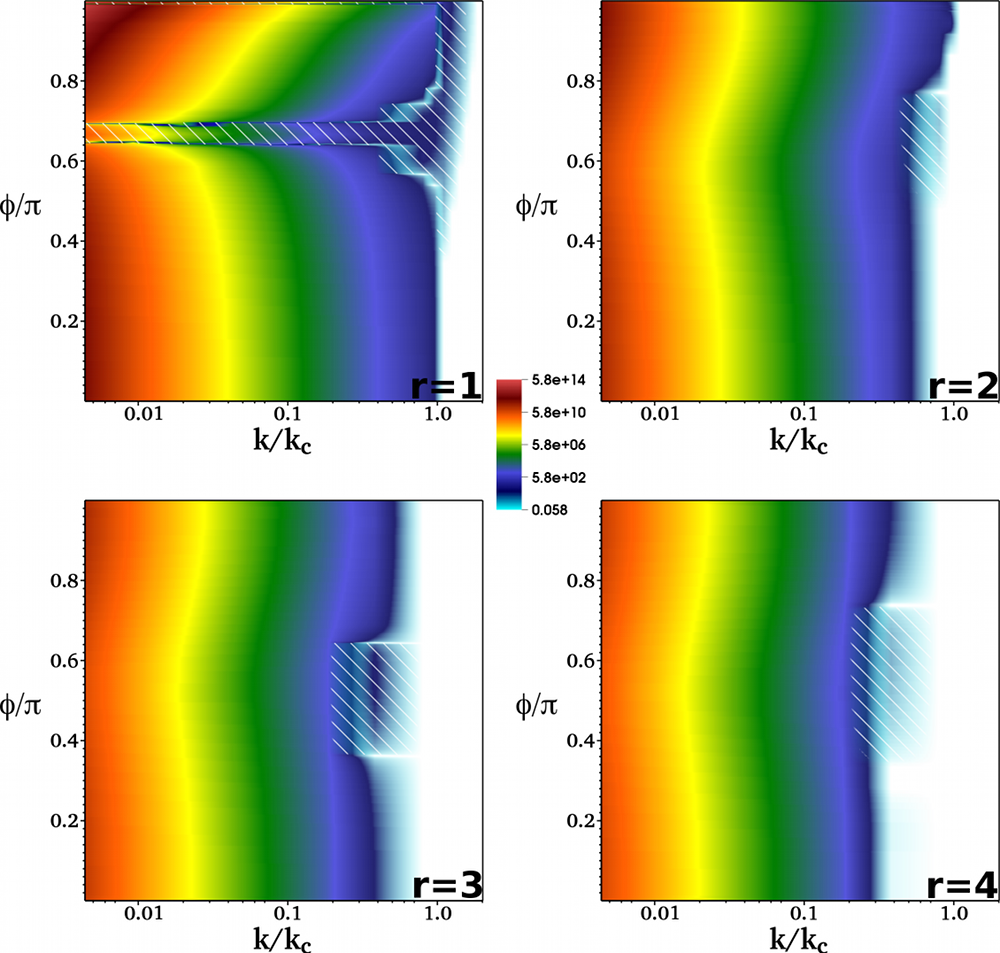}\end{center}
\caption{Slices through $\langle\tau_S^3\rangle$ for $n_B=-5/2$ at constant $r$. Dashed areas are negative.}
\label{PlanesNm25TsTsTs}
\end{figure}

However, the linethroughs at $r=5$ in Figure \ref{LinethroughsNm25}, and the planes in Figure \ref{PlanesNm25TsTsTs} demonstrate that the signal does not remain negative between the equilateral and degenerate lines. Instead, there is a negative trough, or ``river'' running between regions of positivity before a strong negative divergence as $\phi\rightarrow\pi$. Conversely, the small-scale signal is negative for almost all $\phi$. The manner in which this bispectrum changes signs on large scales could potentially lead to very characteristic signatures on the CMB. Furthermore, the fact that the degenerate line possesses a different sign to the scalar signal implies that the \emph{total} magnetic signal -- the sum of all possible CMB bispectra, covering auto- and cross-correlations of all the components of the stress tensor -- will exhibit cancellations.

The planes at $r\in\{2,3,4\}$ demonstrate further oscillations in the signal; a negative region at $k\rightarrow k_c$ grows in area and then declines again with increasing $r$. However, these interesting ``caves'' in the bispectrum are unlikely to leave significant signs on the CMB as the transfer functions and the amplitude of the bispectra close to the damping scale are negligible compared to the largest scales.

In TSS10 the authors also evaluated the result for a ``local isosceles'' shape, which they found to vanish. No such signal is seen here, although naturally some regions of the bispectrum lying between regions of positive and negative value are zero. It is likely that the inclusion of terms they neglected to enable an analytical integration account for the discrepancy.

Evaluating the pivot plane at $k_\star=4.4\times 10^{-3}$, presented in Figure \ref{PivotPlanesNm25}, and the scaling at the equilateral line yields
\be
  \alpha=-4.52\pm 0.12 .
\ee
This value has been recovered sampling only four points -- the levelling of this curve seen in Figure \ref{LinethroughsNm25} contaminates the calculation otherwise. The downside is the increase in the value of the error, which may be mere statistical fluctuation or may be a hint that the integrals should be run at a slightly higher resolution although convergence is still reasonable. Repeating the evaluation along the colinear line (which is very well sampled since there is one less layer of integration) yields
\be
\alpha=-4.4999\pm 0.0003
\ee
which dramatically confirms the scaling to very high precision. $\mathcal{B}(r,\phi)$ shows the pivot plane, plotted on a truncated scale to highlight structure, is non-trivial in nature. There are negative regions at $r=1$ with $\phi=2\pi/3$ and $\phi\rightarrow\pi$; the inset shows a close-up on the region around the degenerate line. The coherence length behaves as for the $\mathcal{B}_{\tau^3}$.

\subsubsection{$\BTtTtTt$}
Finally, consider the $\av{\tau_T^3}$ signal. The colinear, equilateral and degenerate lines in Figure \ref{LinethroughsNm25} show that the colinear signal is identically zero while the equilateral and degenerate lines are of the same order of magnitude as those for $\av{\tau^3}$, with the tensor signal slightly dominating the equilateral line and the scalar signal dominating the degenerate signal. As with the other auto-correlations the degenerate signal is strongly dominant. The signal is positive.

\begin{figure}
\begin{center}\includegraphics[width=0.55\textwidth]{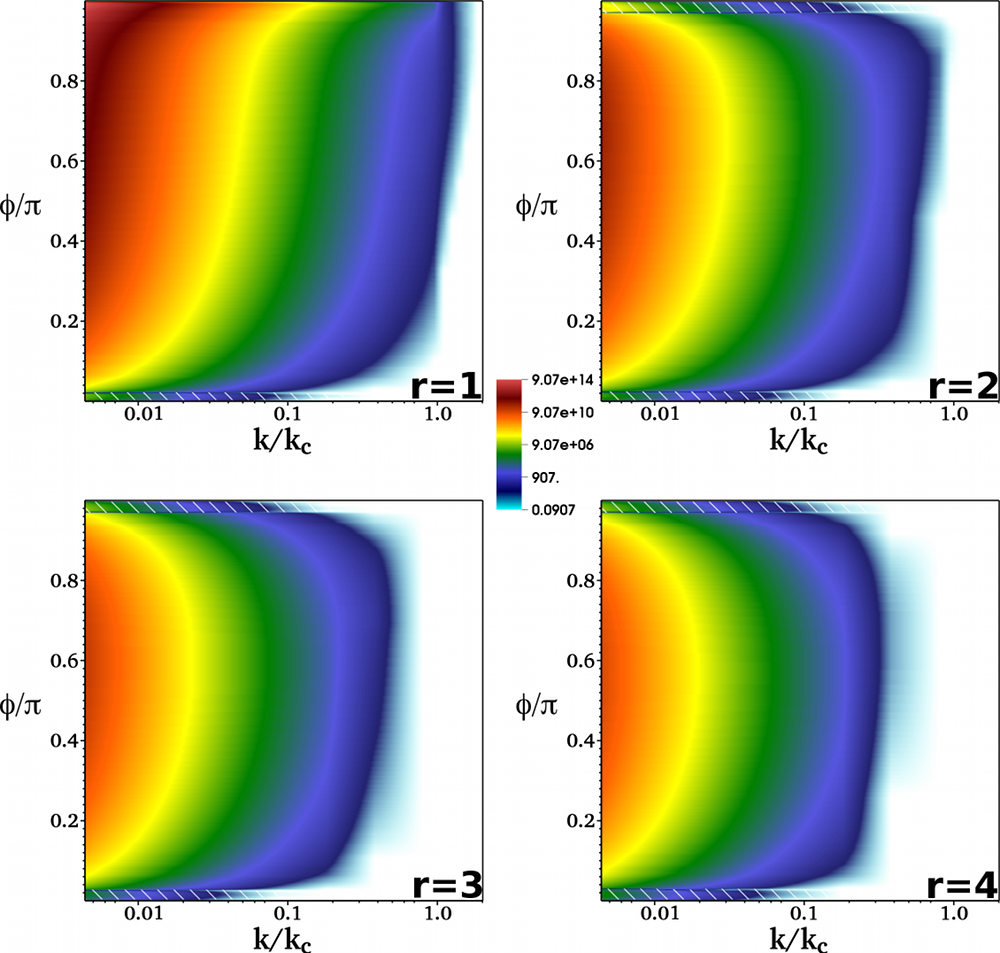}\end{center}
\caption{Slices through $\langle\tau_T^3\rangle$ for $n_B=-5/2$ at constant $r$. Dashed areas are negative.}
\label{PlanesNm25TtTtTt}
\end{figure}

The slices at constant $r$ in Figure \ref{PlanesNm25TtTtTt} reveal a more complicated structure. The colinear plane is identically zero and, by rotation, this implies that away from the degenerate line the squeezed plane is also zero. A negative region lies next to each of the planes. Figure \ref{PivotPlanesNm25} shows the evolution of these regions as $r$ increases at a pivot of $k_\star=4.4\times 10^{-3}$. Otherwise the signal is positive-definite. The scaling recovered along the equilateral line around this plane is found to be
\be
  \alpha=-4.500\pm 0.003 .
\ee
The pivot plane $\mathcal{B}_\star(r,\phi)$ from Figure \ref{PivotPlanesNm25} can then be employed with $\alpha=-9/2$. This plane is, as with most of the other bispectra, non-trivial. The coherence length again behaves as for $\mathcal{B}_{\tau^3}$.

\section{Conclusions}
\label{Conclusions}
This paper presents numerical integrations of the auto-correlations $\BTrTrTr$, $\BTsTsTs$ and $\BTtTtTt$ which are required for an evaluation of the CMB angular bispectrum imprinted by magnetically-induced scalar or tensor modes. On large scales, quantified as lying below a coherence length $k_\mathrm{Coh}=k_\mathrm{Coh}(n_B,r,\phi)$, previous authors have found that the bispectra from ultra-violet fields with $n_B\geq -1$ act as white noise, while those from infra-red fields with $n_B<-1$ scale with $k^{3n_B+3}$, except along the degenerate line where they scale as $q^{2n_B+3}$. This behaviour is confirmed, ultra-violet fields having a coherence length $k_\mathrm{Coh}\sim k_c(\eta)/100$ and infra-red fields having a coherence scale $k\approx k_c(\eta)$, where the values are taken at $r=1$.

In the region applicable to CMB studies the bispectra are well-modelled by a 2-dimensional plane $\mathcal{B}_\star(r,\phi)$ sampled at a pivot of $k=k_\star$. The ultra-violet bispectra typically have power up to a large $r$, with $k_\mathrm{Coh}$ decaying with increasing $r$, so if one were to put an $n_B>-1$ bispectrum onto the CMB in principle one should set a pivot on very large scales and allow it to run with $r$. In practice this would complicate the CMB integration significantly and, while it is not clear that the standard formalism directly applies to such fields, care should be taken. For $n_B<-1$ the evolution of $k_\mathrm{Coh}$ is less important since this remains approximately on the order of $k_\mathrm{Coh}\sim\mathcal{O}(1)$ except in regions when the bispectrum is negligible.

These 2-dimensional planes are non-trivial but (relatively) quick to evaluate since evaluation is only needed at a constant $k_\star$. The sampling can be made arbitrarily fine as $r\rightarrow 1$ and $\phi\rightarrow\pi$ to capture the most important region, and can be relatively sparse outside of these regions. To constrain magnetic parameters from the CMB parameter estimation is required, which involves many multiple evaluations in a Monte-Carlo chain, so this increase in speed is necessary. There are two possibilities: evaluating $\mathcal{B}_\star(r,\phi)$ as requested for each value of $n_B$, or pre-evaluating a number of planes with discrete choices of $n_B$ and interpolating between them as required. The first case would be the ideal but might unfortunately remain prohibitively slow. The latter option rests on the assumption that the plane $\mathcal{B}_\star(r,\phi)$ evolves smoothly with varying $n_B$. In this case one would imagine sampling choices of $n_B\in(-1,-3)$ and stacking these planes into a cube in $\{r,\phi,n_B\}$-space. This issue is under current study.

The CMB signal from the scalar modes is being increasingly studied (see for example SS09, CFPR09, TSS10 and \citet{Cai:2010uw}) and that from the vector auto-correlation has recently also been evaluated \citep{Shiraishi:2010yk,Kahniashvili:2010us}. An immediate consequence of this study is that we can evaluate the CMB angular bispectrum arising from magnetic gravitational waves and we will present the results in a forthcoming work.

Necessary extensions to enable a full parameter estimation include evaluating the scalar cross-correlations $\BTrTrTs$, $\BTrTsTs$, $\BTsTsTs$, the scalar/vector and scalar/tensor cross-correlations $\BTrTvTv$, $\BTsTvTv$, $\BTrTtTt$, $\BTsTtTt$ and the vector/tensor cross-correlation $\BTvTtTv$, the colinear lines of all of which were considered in BC05 and B06. In the first instance we would want to select a constant spectral index $n_B\approx -5/2$ and evaluate the bispectra across a section of the coherent range to confirm the nature of the scalings and the behaviour of permutations of the wavenumbers. For more general $n_B$ it should suffice to evaluate instead the plane $\mathcal{B}_\star(r,\phi)$ at $k_\star\approx 10^{-4}$. We may however find for CMB integration that, given the form of the integration \citep{Ferreira:1998kt,Wang:1999vf,Shiraishi:2010sm} swapping back to $\{k,p,q\}$-space is more convenient. This would depend on the balance between the speed-increase from considering only 2D planes and the complications of remapping from the $\{k,r,\phi\}$ coordinate system, incorporating as well the permutations for cross-correlations, and is currently under study.

Since BC05 and B06 demonstrate that each of these cross-correlations is non-vanishing and of the same order-of-magnitude along the colinear line each one will, in principle, leave traces on the CMB. Furthermore, the fact that the degenerate line possesses a different sign to the scalar signal strongly suggests that the \emph{total} magnetic signal -- the sum of all possible CMB bispectra, covering auto- and cross-correlations of all the components of the stress tensor -- will exhibit cancellations. In particular, since magnetic bispectra are not positive-definite, contributions from unexpected features such as the ``river'' in $\av{\tau_S^3}$, could introduce cancellations in the integrations particularly for indices further from scale-invariance where the degenerate line isn't so dominant. Perhaps more importantly, the complete set ($\BTrTrTr$, $\BTrTrTs$,\ldots,$\BTvTtTv$,\ldots) contributes to the magnetised CMB angular bispectrum and the nature of these terms -- particularly their sign near to the degenerate line -- is unknown. Whether these effects have a significant impact on the CMB bounds is uknown at present, but conceivably they could significantly tighten (or weaken) them. Until the complete set of intrinsic bispectra are known, and then wrapped onto the CMB with the corresponding transfer functions, CMB constraints on magnetic fields from bispectra should be treated with some caution.

\acknowledgments
Many instructive and useful conversations have helped me while I was undertaking this study but I owe a particular debt to Kishore Ananda, Chris Byrnes, Frode Hansen and particularly Richard W. Brown and David Parkinson, who assisted with some of the codes employed in the project, and Robert Crittenden who introduced me to this topic.

\appendix
\section{Appendix A: Angular Terms}
\label{Appendix-AngularTerms}
The angular terms in the bispectra for the scalar auto-correlations are given by
\be
\label{AngularTerm}
  8\FTrTrTr=\betab^2+\gammab^2+\mub^2-\betab\gammab\mub \qquad \mathrm{and} \qquad
  \FTsTsTs=\sum_{n=0}^6\FTsTsTs^{n}
\ee
with
\bea
  -8\FTsTsTs^0&=&9, \nonumber \\
  -8\FTsTsTs^1&=&0, \nonumber \\
  -8\FTsTsTs^2&=&-\Big(
   \betab^2+\gammab^2+\mub^2
   +9(\theta_{kp}^2+\theta_{kq}^2+\theta_{pq}^2)
   +3(\alpha_k^2+\alpha_p^2+\alpha_q^2+\beta_k^2+\beta_p^2+\beta_q^2+\gamma_k^2+\gamma_p^2+\gamma_q^2)
   \Big), \nonumber \\
  -8\FTsTsTs^3&=&3\bigg(
   \mub(\beta_k\gamma_k+\beta_p\gamma_p+\beta_q\gamma_q+\frac{1}{3}\betab\gammab)
   +\gammab(\alpha_k\gamma_k+\alpha_p\gamma_p+\alpha_q\gamma_q)
  +\betab(\alpha_k\beta_k+\alpha_p\beta_p+\alpha_q\beta_q)
    \nonumber \\ && \quad
   +3\theta_{kp}(\alpha_k\alpha_p+\beta_k\beta_p+\gamma_k\gamma_p)
   +3\theta_{kq}(\alpha_k\alpha_q+\beta_k\beta_q+\gamma_k\gamma_q)
   +3\theta_{pq}(\alpha_p\alpha_q+\beta_p\beta_q+\gamma_p\gamma_q)
   +9\theta_{kp}\theta_{kq}\theta_{pq}
   \bigg), \nonumber \\
  -8\FTsTsTs^4&=&-3\bigg(
   \gammab\mub\alpha_k\beta_k+\betab\mub\alpha_p\gamma_p+\betab\gammab\beta_q\gamma_q
   +3\big(\mub\theta_{kp}\beta_k\gamma_p+\gammab\theta_{kq}\alpha_k\gamma_q+\betab\theta_{pq}\alpha_p\beta_q\big)
     \nonumber \\ && \quad
   +3\big(\alpha_k\beta_k(\alpha_p\beta_p+\alpha_q\beta_q)
   +\alpha_p\gamma_p(\alpha_k\gamma_k+\alpha_q\gamma_q)
   +\beta_q\gamma_q(\beta_k\gamma_k+\beta_p\gamma_p)\big)
     \nonumber \\ && \quad
   +9(\theta_{kp}\theta_{kq}\gamma_p\gamma_q+\theta_{kp}\theta_{pq}\beta_k\beta_q+\theta_{kq}\theta_{pq}\alpha_k\alpha_p)
   \bigg),
 \nonumber \\
  -8\FTsTsTs^5&=&9\bigg(
   \mub\alpha_k\beta_k\alpha_p\gamma_p
   +\gammab\alpha_k\beta_k\beta_q\gamma_q+\betab\alpha_p\gamma_p\beta_q\gamma_q
   +3(\theta_{kp}\beta_k\gamma_p\beta_q\gamma_q+\theta_{kq}\alpha_k\alpha_p\gamma_p\gamma_q+\theta_{pq}\alpha_k\beta_k\alpha_p\beta_q)
   \bigg), \nonumber \\
  -8\FTsTsTs^6&=&-27\alpha_k\beta_k\alpha_p\gamma_p\beta_q\gamma_q. \nonumber
\eea
The angular term for the tensor auto-correlation is extremely unwieldy and may be found in B06.

In the colinear limit $\FTsTsTs$ and $\FTtTtTt$ reduce to
\bea
\label{TsTsTs-Colinear}
 8\FTsTsTs&=&9\Bigg(-1+\alpha_k^2+\beta_k^2+\gamma_k^2+\frac{\betab^2+\gammab^2+\mub^2-\betab\gammab\mub}{9}+\frac{\alpha_k\gamma_k\betab\mub+\alpha_k\beta_k\gammab\mub+\beta_k\gamma_k\betab\gammab}{3}
 \nonumber \\ && \qquad\qquad\qquad
-\alpha_k^2\beta_k^2-\alpha_k^2\gamma_k^2-\beta_k^2\gamma_k^2-\alpha_k\beta_k\gamma_k^2\betab-\alpha_k\beta_k^2\gamma_k\gammab-\alpha_k^2\beta_k\gamma_k\mub+3\alpha_k^2\beta_k^2\gamma_k^2\Bigg)
\eea
and
\be
\label{TtTtTt-Colinear}
 \FTtTtTt=0
\ee 
where in the second equation we have used the definitions of the angles given in equation (\ref{Angles-Colinear}).

In the degenerate limit when $\phi=\pi$ the angular terms become
\be
\label{TrTrTr-Degenerate}
  \FTrTrTr=1+\betab^2 ,
\ee
\bea
\label{TsTsTs-Degenerate}
  \lefteqn{\FTsTsTs=1-3\left(\alpha_k^2+\beta_k^2+\gamma_k^2+\beta_k\gamma_k-\alpha_q^2-\beta_q^2\right)+\betab^2
  +3\left(\alpha_k\left(\gamma_k-\beta_k\right)-2\alpha_q\beta_q\right)\betab}
\\ && 
  +9\left(\alpha_k^2\beta_k^2+\alpha_k^2\beta_k\gamma_k+\alpha_k^2\gamma_k^2+\alpha_k\beta_k\alpha_q\beta_q-\alpha_k\gamma_k\alpha_q\beta_q+\beta_k\gamma_k\beta_q^2\right)+3\beta_q^2\betab^2
 -9\alpha_k\left(\beta_k-\gamma_k\right)\beta_q^2\betab
 -27\alpha_k^2\beta_k\gamma_k\beta_q^2
\nonumber
\eea
and
\bea
\label{TtTtTt-Degenerate}
  \FTtTtTt&=&-3
   +3\alpha_k^2+4\beta_k^2+6\beta_k\gamma_k+4\gamma_k^2+3\alpha_q^2+5\beta_q^2+3\betab^2
   -3\left(\alpha_k(\beta_k-\gamma_k)+2\alpha_q\beta_q\right)\betab
   +3\alpha_k\left(\beta_k-\gamma_k\right)\alpha_q\beta_q
\nonumber \\ && \qquad
   +\alpha_k^2\left(\beta_k^2+\gamma_k^2+3\beta_k\gamma_k-\beta_q^2\right)
   +\beta_k^2\left(2\gamma_k^2-2\alpha_q^2-\beta_q^2-\betab^2\right)
   -\gamma_k^2\left(2\alpha_q^2+\beta_q^2+\betab^2\right)
 \\ && \qquad
   -\beta_k\gamma_k\left(\alpha_q^2+2\beta_q^2+2\betab^2\right)
   -\beta_q^2\left(\alpha_q^2+\beta_q^2+\betab^2\right)
   +\alpha_k(\beta_k-\gamma_k)\beta_q^2\betab
  +\left(\beta_k+\gamma_k\right)^2\alpha_q\beta_q\betab
\nonumber \\ && \qquad
   +2\alpha_q\beta_q^3\betab
   +\alpha_k^2\left(2\beta_k\gamma_k\left(\beta_k\gamma_k-\beta_q^2\right)+\beta_q^4\right)
  -\alpha_k\left(\beta_k-\gamma_k\right)\alpha_q\beta_q^3
   -\beta_k\gamma_k\alpha_q^2\beta_q^2 .
\nonumber
\eea

\section[toc=UV Bispectra on Large-Scales]{Appendix B: Ultra-Violet Degenerate Bispectra on Large Scales}
\label{AppendixDegenerate}
Analytic solutions can be found for the degenerate bispectra across the entire range of $k$ for $n_B\geq -1$ and are available from the author on request or on his website. For simplicity only the Laurent expansions around $k=0$ are presented here, representing the degenerate bispectra on the largest scales. Large-scale solutions in this r\'egime can also be found for the colinear bispectra and the general case by reducing the angular terms to the large-scale limits, which depend only on $\alpha_k$; in these cases the expansions here are accurate only to linear order.

In the infra-red r\'egime when $n_B<-1$ the solutions include a term dependant on the logarithm of an infra-red cut-off scale. For example, on large scales when $n_B=-3/2$ we find
\be
  \BTrTrTr(\mathrm{\phi=\pi})=\left(\left(\frac{16}{9}+\frac{32}{3}\ln(2)-\frac{8}{3}\pi+\frac{16}{3}\ln(k/k_\mathrm{min})\right)k^{-3/2}-\frac{16}{3}-2k-2k^2\right)\pi k_c^3
\ee
where $k_\mathrm{min}$ is the infra-red cut-off. Since there is no immediate model motivating such an infra-red cut-off (barring a realisation on a finite grid) these solutions are not presented here, although they are straightforward to recover.

The results presented in this section complement those in Appendix E.2 of B06 and the more general but colinear results of CFPR09, extending them to the degenerate bispectrum and the anisotropic and tensor auto-correlations. In both cases we extend the solutions to a wider range of $n_B$. Comparison with the full solutions for $n_B>-1$ demonstrates that these expansions (to third order in $k$) are valid for $k\lesssim k_c(\eta)/20$.

When $n_B=3$,
\be
\begin{array}{c}
  \BTrTrTr=\left(\frac{2}{3}-2k+\frac{18}{5}k^2-\frac{13}{3}k^3\right)\pi k_c^3, \qquad
  \BTsTsTs=\left(\frac{34}{105}-\frac{1}{8}k-\frac{62}{75}k^2+\frac{331}{240}k^3\right)\pi k_c^3, \\
  \BTtTtTt=\left(\frac{68}{315}-\frac{19}{24}k+\frac{884}{525}k^2-\frac{107}{45}k^3\right)\pi k_c^3 .
\end{array}
\ee

When $n_B=5/2$,
\be
\begin{array}{c}
  \BTrTrTr=\left(\frac{16}{21}-2k+\frac{139}{51}k^2-\frac{59}{24}k^3\right)\pi k_c^3, \qquad
  \BTsTsTs=\left(\frac{272}{735}-\frac{1}{8}k-\frac{290}{357}k^2+\frac{509}{480}k^3\right)\pi k_c^3, \\
  \BTtTtTt=\left(\frac{544}{2,205}-\frac{19}{24}k+\frac{852}{595}k^2-\frac{4,781}{2,880}k^3\right)\pi k_c^3 .
\end{array}
\ee

When $n_B=2$,
\be
\begin{array}{c}
  \BTrTrTr=\left(\frac{8}{9}-2k+\frac{44}{21}k^2-\frac{4}{3}k^3\right)\pi k_c^3, \qquad
  \BTsTsTs=\left(\frac{136}{315}-\frac{1}{8}k-\frac{86}{105}k^2+\frac{19}{24}k^3\right)\pi k_c^3, \\
  \BTtTtTt=\left(\frac{272}{945}-\frac{19}{24}k+\frac{124}{105}k^2-\frac{77}{72}k^3\right)\pi k_c^3 .
\end{array}
\ee

When $n_B=3/2$,
\be
\begin{array}{c}
  \BTrTrTr=\left(\frac{16}{15}-2k+\frac{47}{33}k^2-\frac{11}{24}k^3\right)\pi k_c^3, \qquad
  \BTsTsTs=\left(\frac{272}{525}-\frac{1}{8}k-\frac{998}{1,155}k^2+\frac{55}{96}k^3\right)\pi k_c^3, \\
  \BTtTtTt=\left(\frac{544}{1,575}-\frac{19}{24}k+\frac{1,076}{1,155}k^2-\frac{349}{576}k^3\right)\pi k_c^3 .
\end{array}
\ee

When $n_B=1$,
\be
\begin{array}{c}
  \BTrTrTr=\left(\frac{4}{3}-2k+\frac{2}{3}k^2+\frac{1}{6}k^3\right)\pi k_c^3, \qquad
  \BTsTsTs=\left(\frac{68}{105}-\frac{1}{8}k-\frac{104}{105}k^2+\frac{97}{240}k^3\right)\pi k_c^3, \\
  \BTtTtTt=\left(\frac{136}{315}-\frac{19}{24}k+\frac{24}{35}k^2-\frac{97}{360}k^3\right)\pi k_c^3 .
\end{array}
\ee

When $n_B=1/2$,
\be
\begin{array}{c}
  \BTrTrTr=\left(\frac{16}{9}-2k-\frac{1}{3}k^2+\frac{13}{24}k^3\right)\pi k_c^3, \qquad
  \BTsTsTs=\left(\frac{272}{315}-\frac{1}{8}k-\frac{706}{525}k^2+\frac{137}{480}k^3\right)\pi k_c^3, \\
  \BTtTtTt=\left(\frac{544}{945}-\frac{19}{24}k+\frac{236}{525}k^2-\frac{173}{2,880}k^3\right)\pi k_c^3 .
\end{array}
\ee

When $n_B=0$,
\be
\begin{array}{c}
  \BTrTrTr=\left(\frac{8}{3}-2k-\frac{8}{3}k^2+\left(\frac{1}{4}\pi^2+\frac{2}{3}\right)k^3\right)\pi k_c^3, \qquad
  \BTsTsTs=\left(\frac{136}{105}-\frac{1}{8}k-\frac{62}{21}k^2+\left(\frac{13}{60}+\frac{13}{64}\pi^2\right)k^3\right)\pi k_c^3, \\
  \BTtTtTt=\left(\frac{272}{315}-\frac{19}{24}k+\frac{4}{15}k^2+\left(\frac{1}{45}-\frac{1}{64}\pi^2\right)k^3\right)\pi k_c^3 .
\end{array}
\ee

When $n_B=-1/2$,
\be
\begin{array}{c}
  \BTrTrTr=\left(\frac{16}{3}-2k-\frac{128}{15}k^{3/2}+\frac{17}{3}k^2+\frac{13}{24}k^3\right)\pi k_c^3, \qquad
  \BTsTsTs=\left(\frac{272}{105}-\frac{1}{8}k-\frac{79,888}{9,945}k^{3/2}+\frac{82}{15}k^2+\frac{19}{96}k^3\right)\pi k_c^3, \\
  \BTtTtTt=\left(\frac{544}{315}-\frac{19}{24}k+\frac{608}{29,835}k^{3/2}-\frac{12}{35}k^2-\frac{13}{576}k^3\right)\pi k_c^3 .
\end{array}
\ee

Finally, when $n_B=-1$,
\be
\begin{array}{c}
  \BTrTrTr=\left(4-8\ln(k)-2k+\frac{2}{3}k^2+\frac{1}{6}k^3\right)\pi k_c^3, \qquad
  \BTsTsTs=\left(-\frac{2,428}{1,225}-\frac{136}{35}\ln(k)-\frac{1}{8}k+\frac{142}{105}k^2+\frac{11}{48}k^3\right)\pi k_c^3, \\
  \BTtTtTt=\left(\frac{39,472}{11,025}-\frac{272}{105}\ln(k)-\frac{19}{24}k-\frac{44}{105}k^2-\frac{7}{36}k^3\right)\pi k_c^3 .
\end{array}
\ee

\bibliographystyle{apj}
\bibliography{IntrinsicBispectraOfCosmicMagneticFields}

\end{document}